\documentclass{imamci}

\usepackage{natbib}
\usepackage{xcolor}
\usepackage{graphicx}
\usepackage{amsmath,amsthm,bm,mathrsfs}

 \newtheoremstyle{theorem}{6pt}{6pt}{\rm}{}{\sffamily}{ }{ }{}
 \theoremstyle{theorem}

 \newtheoremstyle{algorithm}{6pt}{6pt}{\rm}{}{\sffamily}{ }{ }{}
 \theoremstyle{algorithm}

 \newtheoremstyle{lemma}{6pt}{6pt}{\rm}{}{\sffamily}{ }{ }{}
 \theoremstyle{lemma}

\newtheoremstyle{case}{6pt}{6pt}{\rm}{}{\sffamily}{. }{ }{}
 \theoremstyle{case}

 \newtheoremstyle{statement}{6pt}{6pt}{\rm}{}{\sffamily}{ }{ }{}
\theoremstyle{statement}

 \newtheoremstyle{corollary}{6pt}{6pt}{\rm}{}{\sffamily}{ }{ }{}
 \theoremstyle{corollary}

  \newtheoremstyle{definition}{6pt}{6pt}{\rm}{}{\sffamily}{ }{ }{}
 \theoremstyle{definition}

\newtheoremstyle{example}{6pt}{6pt}{\rm}{}{\sffamily}{ }{ }{}
\theoremstyle{example}

\newtheoremstyle{remark}{6pt}{6pt}{\rm}{}{\sffamily}{ }{ }{}
\theoremstyle{remark}

\newtheoremstyle{approximation}{6pt}{6pt}{\rm}{}{\sffamily}{ }{ }{}
\theoremstyle{approximation}

\newtheoremstyle{scheme}{6pt}{6pt}{\rm}{}{\sffamily}{ }{ }{}
\theoremstyle{scheme}

\newtheoremstyle{Algorithm}{6pt}{6pt}{\rm}{}{\sffamily}{ }{ }{}
\theoremstyle{Algorithm}

\newtheoremstyle{Assumption}{6pt}{6pt}{\rm}{}{\sffamily}{ }{ }{}
\theoremstyle{Assumption}

\newtheoremstyle{proposition}{6pt}{6pt}{\rm}{}{\sffamily}{ }{ }{}
\theoremstyle{proposition}

\newtheoremstyle{hypo}{6pt}{6pt}{\rm}{}{\sffamily}{ }{ }{}
 \theoremstyle{hypo}

  \newtheoremstyle{Step}{6pt}{6pt}{\rm}{}{}{ }{ }{}
 \theoremstyle{Step}

\numberwithin{equation}{section}

\begin{document}

%%%%%%%%%%%%%%%%%
\title{Frequency-weighted $\mathcal{H}_2$-pseudo-optimal model order reduction}
\author{ {\sc Umair~Zulfiqar$^{*}$ and Victor~Sreeram}\\[2pt]
School of Electrical, Electronics and Computer Engineering, The University of Western Australia (UWA), Perth Australia\\
$^*$\textnormal{Corresponding author. Email: umair.zulfiqar@research.uwa.edu.au}\\
\textnormal{victor.sreeram@uwa.edu.au}\\[4pt]
{\sc Mian~Ilyas~Ahmad}\\[2pt]
 Research Centre for Modelling and Simulation, National University of Sciences and Technology (NUST), Islamabad Pakistan\\
\textnormal{m.ilyas@rcms.nust.edu.pk}\\
{\sc and}\\[4pt]
{\sc  Xin~Du}\\[2pt]
School of Mechatronic Engineering and Automation,\\
and Shanghai Key Laboratory of Power Station Automation Technology,\\
Shanghai University, Shanghai, China\\
\textnormal{duxin@shu.edu.cn}
\vspace*{6pt}}
\pagestyle{headings}
\markboth{U. ZULFIQAR ET AL.}{\rm FREQUENCY-WEIGHTED $\mathcal{H}_2$-PSEUDO-OPTIMAL MODEL ORDER REDUCTION}
\maketitle

%%%%%%%%%%%%%%%%%abstract style
%Two grouping braces are necessary in abstract environment
%first argument contains abstract text; second argument contains keywords
%text

\begin{abstract}
{The frequency-weighted model order reduction techniques are used to find a lower-order approximation of the high-order system that exhibits high-fidelity within the frequency region emphasized by the frequency weights. In this paper, we investigate the frequency-weighted $\mathcal{H}_2$-pseudo-optimal model order reduction problem wherein a subset of the optimality conditions for the local optimum is attempted to be satisfied. We propose two iteration-free algorithms, for the single-sided frequency-weighted case of $\mathcal{H}_2$-model reduction, where a subset of the optimality conditions is ensured by the reduced system. In addition, the reduced systems retain the stability property of the original system. We also present an iterative algorithm for the double-sided frequency-weighted case, which constructs a reduced-order model that tends to satisfy a subset of the first-order optimality conditions for the local optimum. The proposed algorithm is computationally efficient as compared to the existing algorithms. We validate the theory developed in this paper on three numerical examples.}
{$\mathcal{H}_2$-optimal, frequency-weighted, model reduction, pseudo-optimal.}
\end{abstract}
%%%%%%%%%%%%%%%%%%%%%%%%%%%%%%%

%%%%%%%%%%%%%%section A%%%%%%%%%
\section{Introduction}
The process of obtaining a reduced-order approximation of the original high-order system that can effectively act as a surrogate in the design and analysis is called model order reduction (MOR). The quality of the approximation is evaluated by using error expressions in some defined norm. Various measures to quantify the error leads to different MOR problems. In frequency-weighted MOR (FWMOR), it is aimed to construct a ROM of the original high-order model that ensures less frequency-weighted approximation error. The frequency region, wherein a superior accuracy is desired, is emphasized by the frequency weights.

Controller reduction is an important application of FWMOR techniques. The optimal controller design procedures like linear-quadratic-Gaussian (LQG), $\mathcal{H}_2$, and $\mathcal{H}_\infty$ controllers are theoretically well-grounded in guaranteeing the desired closed-loop performance. However, their existence and implementation for large-scale plants are the real challenges. The control law in these analytical procedures is commonly computed by solving linear matrix equations (Riccati equations) which encounter several numerical difficulties like ill-conditioning, Hamiltonian too close to the imaginary axis, and excessive memory requirements as the size of the plants becomes high. Moreover, the controller obtained is of the same order as that of the plant, which restricts its practical implementation for high-order plants. The MOR reduction algorithms that are used to obtain a reduced-order controller for the original high-order plant tend to preserve the closed-loop characteristics. This necessitates the inclusion of both the plant and controller in the approximation criterion of the MOR algorithms, which mostly makes this problem a FWMOR criterion \citep{enns1985,rivera1987control,anderson1989controller,mcfarlane1990reduced,obinata2012model}. Several algorithms have been developed since the late 1980s for this purpose, and several closed-loop approximation criteria have been considered so far like closed-loop stability, closeness of the closed-loop transfer function, and controller input spectrum; see \citep{obinata2012model} for a detailed survey.

Balanced truncation (BT) is an important MOR technique for which several algorithms and extensions have been presented over the last three decades \citep{moore1981principal}. It is known for its good accuracy, stability preservation, and error bound \citep{enns1984model}. However, BT can only be applied to models of moderate size due to its excessive computational cost. There exist some generalizations to extend its applicability to large-scale systems by replacing the large-scale Lyapunov equations with their low-rank approximations \citep{mehrmann2005balanced}. Enns generalized BT to frequency-weighted BT (FWBT) for obtaining a superior accuracy within the frequency region emphasized by the frequency weights \citep{enns1984model,enns1985}. Several other extensions of FWBT are also reported in the literature, like \citep{sreeram1995new,sreeram1995frequency,wang1999new,sreeram2002properties,varga2003accuracy}, which are surveyed in \citep{ghafoor2008survey}. The optimal solution for the FWMOR problem in the $\mathcal{H}_\infty$-norm is hard to find, and the available algorithms based on the solution of large-scale linear matrix inequalities (LMIs) can only be applied to small-scale systems \citep{wu2003computationally}. In \citep{halevi1992frequency}, the optimal solution for the FWMOR problem in the $\mathcal{H}_2$-norm is considered, and the first-order optimality conditions for the local optimum are derived. The optimization algorithm given to enforce these Gramian based-conditions becomes computationally infeasible for even moderate size systems. The $\mathcal{H}_2$-optimal frequency-weighted MOR has also been attempted in \citep{valentin1997lmi,yan1997convergent,huang2001new}.

A popular approach to obtain an optimal ROM for the FWMOR problem in the Hankel-norm is the frequency-weighted Hankel norm approximation (FWOHNA) \citep{latham1985frequency,anderson1986weighted,hung1986optimal,zhou1995frequency}. Several extensions of FWOHNA are also reported in the literature to guarantee the stability of the ROM \citep{wang1998new,kumar2013improved,kumar2020factorization}.

Moment matching is another important class of MOR techniques wherein the transfer function of the original system and some of its derivatives are matched by the reduced order system at some selected interpolation points. Unlike BT and its generalizations, moment matching techniques are computationally efficient due to their Krylov subspace-based implementation \citep{benner2005dimension}. The solution for the $\mathcal{H}_2$-optimal MOR problem can be found in a computationally efficient way using Krylov subspace-based moment matching algorithm, i.e., the iterative rational Krylov algorithm (IRKA) \citep{gugercin2008h_2}. It is shown in \citep{gugercin2008h_2} that a ROM is a local optimum for the $\mathcal{H}_2$-optimal MOR problem if it interpolates the original system at the mirror images of its poles. The poles of the ROM that satisfies such a condition are not known a priori. Thus an iterative algorithm is presented in \citep{gugercin2008h_2}, which updates the interpolation points as the mirror images of the poles after each iteration until it converges to some fixed points. IRKA is extended to multi-input multi-output (MIMO) systems in \citep{van2008h2}. IRKA normally converges quickly even if it is initialized randomly; however, the convergence is considerably slow as the number of inputs and outputs increases. In \citep{wolf2014h}, an iteration-free algorithm is proposed, i.e., Pseudo-optimal Rational Krylov (PORK) algorithm that constructs a ROM in a single-run, which satisfies a subset of the optimality conditions. PORK is based on the parametrization of the ROM approach of \citep{astolfi2010model,ilyasphd}. It is shown in \citep{astolfi2010model,ilyasphd} that a ROM generated by the rational Krylov subspace methods can be parameterized to enforce some desired properties without effecting the interpolation condition.

IRKA is also extended heuristically to the FWMOR scenario in \citep{anic2013interpolatory} for single-input single-output (SISO) systems. The algorithm in \citep{anic2013interpolatory} generates a ROM that offers less weighted $\mathcal{H}_2$-norm error; however, it does not satisfy any optimality conditions. In \citep{breiten2015near}, the Gramian-based optimality conditions for the frequency-weighted $\mathcal{H}_2$-MOR problem (derived in \citep{halevi1992frequency}) are expressed as IRKA-like interpolation conditions. Then an IRKA-like algorithm is presented, which generates a ROM that nearly satisfies these optimality conditions. The algorithm is thus named as near-optimal weighted interpolation (NOWI). Similar to IRKA, it is an iterative algorithm wherein the interpolation data is updated after each iteration. It avoids the computation of two large-scale Lyapunov equations required in FWBT \citep{enns1984model} and large-scale linear matrix equations in \citep{halevi1992frequency}. Thus, it is computationally less expensive than FWBT and \citep{halevi1992frequency}.

In this paper, we first present an algorithm for the input frequency-weighted case of $\mathcal{H}_2$-optimal MOR, which satisfies the interpolation and optimality conditions that NOWI fails to satisfy. Our approach is inspired by PORK, and we parameterize the ROM to enforce a subset of first-order optimality conditions. Note that NOWI is an iterative algorithm with no assurance of convergence. Unlike NOWI, the proposed algorithm is iteration-free, and it constructs the ROM in a single-run. Its computational cost is fraction of that of NOWI, and unlike NOWI, it also preserves the stability of the original system in the ROM. We then present the output frequency-weighted case of $\mathcal{H}_2$-optimal MOR, which is the dual of the first case. It also satisfies a subset of first-order optimality conditions, and the stability of the original system is also preserved in the ROM. These two algorithms also generate approximate frequency-weighted Gramians at a significantly reduced computational cost, which can be used to perform an approximate FWBT. We also present an IRKA-like iterative algorithm for the double-sided case of the frequency-weighted $\mathcal{H}_2$-optimal MOR, which is computationally more efficient than FWBT. We test our algorithms on benchmark numerical examples to validate the theory presented in the paper.
\section{Preliminaries}
Let $G(s)$, $\mathscr{V}(s)$, and $\mathscr{W}(s)$ be the original system, input frequency weight, and the output frequency weight, respectively, i.e.,
\begin{align}
G(s)=&C(sI-A)^{-1}B+D,\nonumber\\
\mathscr{V}(s)=&\mathscr{C}_v(sI-\mathscr{A}_v)^{-1}\mathscr{B}_v+\mathscr{D}_v,\nonumber\\
\mathscr{W}(s)=&\mathscr{C}_w(sI-\mathscr{A}_w)^{-1}\mathscr{B}_w+\mathscr{D}_w,\nonumber
\end{align} where
$A\in\mathbb{R}^{n\times n}$, $B\in\mathbb{R}^{n\times m}$, $C\in\mathbb{R}^{p\times n}$, $D\in\mathbb{R}^{p\times m}$, $\mathscr{A}_v\in\mathbb{R}^{n_v\times n_v}$, $\mathscr{B}_v\in\mathbb{R}^{n_v\times m}$, $\mathscr{C}_v\in\mathbb{R}^{m\times n_v}$, $\mathscr{D}_v\in\mathbb{R}^{m\times m}$, $\mathscr{A}_w\in\mathbb{R}^{n_w\times n_w}$, $\mathscr{B}_w\in\mathbb{R}^{n_w\times p}$, $\mathscr{C}_w\in\mathbb{R}^{p\times n_w}$, and $\mathscr{D}_w\in\mathbb{R}^{p\times p}$. Further, suppose that $G(s)$, $\mathscr{V}(s)$, and $\mathscr{W}(s)$ are stable.

The FWMOR problem is to construct a $r^{th}$-order ROM $\tilde{G}_r(s)$ such that the weighted error $||\mathscr{W}(s)\big(G(s)-\tilde{G}_r(s)\big)\mathscr{V}(s)||$ is small. The original model is projected onto a reduced subspace to obtain a ROM as the following
\begin{align}
\tilde{G}_r(s)&=C\tilde{V}_r(sI-\tilde{W}_r^TA\tilde{V}_r)^{-1}\tilde{W}_r^TB+\tilde{D}_r\nonumber\\
&=\tilde{C}_r(sI-\tilde{A}_r)^{-1}\tilde{B}_r+\tilde{D}_r\nonumber
\end{align} where $\tilde{V}_r\in\mathbb{R}^{n\times r}$ and $\tilde{W}_r\in\mathbb{R}^{n\times r}$ are the input and output reduction subspaces.

The controllability Gramians of the pairs $(A,B)$, $(\tilde{A}_r,\tilde{B}_r)$, and $(\mathscr{A}_v,\mathscr{B}_v)$ are the solutions of following Lyapunov equations
\begin{align}
AP+PA^T+BB^T=0,\nonumber\\
\tilde{A}_r\tilde{P}+\tilde{P}\tilde{A}_r^T+\tilde{B}_r\tilde{B}_r^T=0,\nonumber\\
\mathscr{A}_v\mathscr{P}_v+\mathscr{P}_v\mathscr{A}_v^T+\mathscr{B}_v\mathscr{B}_v^T=0.\label{7}
\end{align}

The observability Gramians of the pairs $(A,C)$, $(\tilde{A}_r,\tilde{C}_r)$, and $(\mathscr{A}_w,\mathscr{C}_w)$ are the solutions of following Lyapunov equations
\begin{align}
A^TQ+QA+C^TC=0,\nonumber\\
\tilde{A}_r^T\tilde{Q}+\tilde{Q}\tilde{A}_r+\tilde{C}_r^T\tilde{C}_r=0,\nonumber\\
\mathscr{A}_w^T\mathscr{Q}_w+\mathscr{Q}_w\mathscr{A}_w+\mathscr{C}_w^T\mathscr{C}_w=0.\label{10}
\end{align}
\subsection{FWBT \citep{enns1984model}}
A state-space realization is called a balanced realization if its states are equally controllable and observable. The controllability and observability Gramians of such a realization are equal and diagonal whose diagonal entries correspond to the square of Hankel singular values. In other words, the states are arranged according to the Hankel singular values. In BT, the states associated with the least Hankel singular values are truncated due to their negligible share in the overall energy transfer. Enns proposed a frequency-weighted generalization of BT wherein a frequency-weighted balanced realization is first sought, and then the states associated with the least frequency-weighted Hankel singular values are truncated. The ROM thus obtained generally ensures that the weighted-error $||\mathscr{W}(s)\big(G(s)-\tilde{G}_r(s)\big)\mathscr{V}(s)||$ is small.

Let $\mathscr{H}_i(s)$ be the input augmented system with the following state-space realization
\begin{align}
\mathscr{H}_i(s)=G(s)\mathscr{V}(s)=\mathscr{C}_i(sI-\mathscr{A}_i)^{-1}\mathscr{B}_i+\mathscr{D}_i\nonumber
\end{align} where
\begin{align}
\mathscr{A}_i=&\begin{bmatrix}A&B\mathscr{C}_v\\0&\mathscr{A}_v\end{bmatrix}, \hspace*{0.5cm}\mathscr{B}_i=\begin{bmatrix}B\mathscr{D}_v\\\mathscr{B}_v\end{bmatrix},\label{13F}\\
\mathscr{C}_i=&\begin{bmatrix}C &D\mathscr{C}_v\end{bmatrix}, \hspace*{0.5cm}\mathscr{D}_i=D\mathscr{D}_v.
\end{align}
The controllability Gramian $\mathscr{P}_i$ of the pair $(\mathscr{A}_i,\mathscr{B}_i)$ is the solution of the following Lyapunov equation
\begin{align}
\mathscr{A}_i\mathscr{P}_i+\mathscr{P}_i\mathscr{A}_i^T+\mathscr{B}_i\mathscr{B}_i^T=0.\label{15C}
\end{align}
$\mathscr{P}_i$ can be partitioned according to $\mathscr{A}_i$ as
\begin{align}
\mathscr{P}_i=\begin{bmatrix}P_e&\mathscr{P}_{12}\\\mathscr{P}_{12}^T&\mathscr{P}_v\end{bmatrix}.\label{16C}
\end{align} Let us define $B_1$ and $B_2$ as the following \begin{align}
B_{1}&=\begin{bmatrix}B&\mathscr{P}_{12}\mathscr{C}_v^T&B\mathscr{D}_v\end{bmatrix}\textnormal{ and }B_2=\begin{bmatrix}\mathscr{P}_{12}\mathscr{C}_v^T&B&B\mathscr{D}_v\end{bmatrix}.\nonumber
\end{align}
Then it can be readily verified by expanding the equation (\ref{15C}) according to (\ref{16C}) that $P_e$ and $\mathscr{P}_{12}$ solve the following matrix equations
\begin{align}
AP_{e}+P_{e}A^T+B_1B_2^T=0,\label{17C}\\
A\mathscr{P}_{12}+\mathscr{P}_{12}\mathscr{A}_v^T+B(\mathscr{C}_v\mathscr{P}_v+\mathscr{D}_v\mathscr{B}_v^T)=0.\label{18C}
\end{align}
$P_e$ is called the frequency-weighted controllability Gramian of the pair $(A,B)$.

Let $\mathscr{H}_o(s)$ be the output augmented system with the following state-space realization
\begin{align}
\mathscr{H}_o(s)=\mathscr{W}(s)G(s)=\mathscr{C}_o(sI-\mathscr{A}_o)^{-1}\mathscr{B}_o+\mathscr{D}_o\nonumber
\end{align} where
\begin{align}
\mathscr{A}_o=&\begin{bmatrix}A&0\\\mathscr{B}_wC&\mathscr{A}_w\end{bmatrix}, \hspace*{0.5cm}\mathscr{B}_o=\begin{bmatrix}B\\\mathscr{B}_wD\end{bmatrix},\label{17F}\\
\mathscr{C}_o=&\begin{bmatrix}\mathscr{D}_wC &\mathscr{C}_w\end{bmatrix}, \hspace*{0.5cm}\mathscr{D}_o=\mathscr{D}_wD.
\end{align}
The observability Gramian $\mathscr{Q}_o$ of the pair $(\mathscr{A}_o,\mathscr{C}_o)$ is the solution of the following Lyapunov equation
\begin{align}
\mathscr{A}_o^T\mathscr{Q}_o+\mathscr{Q}_o\mathscr{A}_o+\mathscr{C}_o^T\mathscr{C}_o=0.\label{21C}
\end{align}
$\mathscr{Q}_o$ can be partitioned according to $\mathscr{A}_o$ as
\begin{align}
\mathscr{Q}_o=\begin{bmatrix}Q_e&\mathscr{Q}_{12}\\\mathscr{Q}_{12}^T&\mathscr{Q}_w\end{bmatrix}.\label{22C}
\end{align} Let us define $C_1$ and $C_2$ as the following
\begin{align}
C_{1}^T=\begin{bmatrix}C^T&\mathscr{Q}_{12}\mathscr{B}_w&C^T\mathscr{D}_w^T\end{bmatrix}\textnormal{ and }C_{2}^T=\begin{bmatrix}\mathscr{Q}_{12}\mathscr{B}_w&C^T&C^T\mathscr{D}_w^T\end{bmatrix}.\nonumber
\end{align}
Then it can be readily verified by expanding the equation (\ref{21C}) according to (\ref{22C}) that $Q_e$ and $\mathscr{Q}_{12}$ are the solutions of following matrix equations
\begin{align}
A^TQ_e+Q_eA+C_{1}^TC_{2}=0\label{25C},\\
A^T\mathscr{Q}_{12}+\mathscr{Q}_{12}\mathscr{A}_w+C^T(\mathscr{B}_w^T\mathscr{Q}_w+\mathscr{D}_w^T\mathscr{C}_w)=0.\label{24F}
\end{align}
$Q_e$ is called the frequency-weighted observability Gramian of the pair $(A,C)$.

Let $(A_t,B_t,C_t,D_t)$ be the state-space realization of $G(s)$ obtained by the applying the similarity transformation $T$, i.e.,
\begin{align}
A_t&=T^{-1}AT,&& B_t=T^{-1}B,\nonumber\\
C_t&=CT,&& D_t=D.\nonumber
\end{align} Then the frequency-weighted controllability Gramian $P_{t,e}$ and the frequency-weighted observability Gramian $Q_{t,e}$ of $(A_t,B_t,C_t)$ are related with $P_e$ and $Q_e$ as
\begin{align}
P_{t,e}=T^{-1}P_eT^{-T}\textnormal{ and } Q_{t,e}=T^TQ_eT.\nonumber
\end{align}

A frequency-weighted balanced realization is the one that has $P_{t,e}=Q_{t,e}=diag(\bar{\sigma_i},\cdots,\bar{\sigma}_n)$ where $\bar{\sigma_i}\geq \bar{\sigma_{i+1}}$ and $\bar{\sigma_i}$ are the frequency-weighted Hankel singular values. The balancing transformation is thus obtained as a contragradient transformation matrix of $P_e$ and $Q_e$, which arranges the states in descending order of their frequency-weighted Hankel singular values. The states associated with the negligible values of $\bar{\sigma_i}$ are truncated to obtain a ROM. Thus $\tilde{V}_r$ and $\tilde{W}_r$ in FWBT are obtained as $\tilde{V}_r=TZ^T$ and $\tilde{W}_r=T^{-1}Z^T$ where $Z=\begin{bmatrix}I_{r\times r}& 0_{r\times (n-r)}\end{bmatrix}$ and $T^{-1}P_eT^{-T}=T^TQ_eT=diag(\bar{\sigma_i},\cdots,\bar{\sigma}_n)$. The ROM is computed as
\begin{align}
\tilde{G}_r(s)=C\tilde{V}_r(sI-\tilde{W}_r^TA\tilde{V}_r)^{-1}\tilde{W}_r^TB+D.\nonumber
\end{align}
\subsection{NOWI \citep{breiten2015near}}
In \citep{breiten2015near}, the single-side case of the FWMOR is considered, i.e., $\mathscr{W}(s)=I$, and the first-order optimality conditions for the problem $||\big(G(s)-\tilde{G}_r(s)\big)\mathscr{V}(s)||^2_{\mathcal{H}_2}$ are formulated in terms of bi-tangential Hermite interpolation conditions. Then an iterative framework is presented, which constructs a ROM that nearly satisfies these interpolation conditions. We first define all the necessary variables required to express the optimality conditions.

Let the input augmented system $\tilde{\mathscr{H}}_i(s)$ has the following state-space realization
\begin{align}
\tilde{\mathscr{H}}_i(s)=\tilde{G}_r(s)\mathscr{V}(s)=\tilde{\mathscr{C}}_i(sI-\tilde{\mathscr{A}}_i)^{-1}\tilde{\mathscr{B}}_i+\tilde{\mathscr{D}}_i\nonumber
\end{align} where
\begin{align}
\tilde{\mathscr{A}}_i=&\begin{bmatrix}\tilde{A}_r&\tilde{B}_r\mathscr{C}_v\\0&\mathscr{A}_v\end{bmatrix}, \hspace*{0.5cm}\tilde{\mathscr{B}}_i=\begin{bmatrix}\tilde{B}_r\mathscr{D}_v\\\mathscr{B}_v\end{bmatrix},\nonumber\\
\tilde{\mathscr{C}}_i=&\begin{bmatrix}\tilde{C}_r &\tilde{D}_r\mathscr{C}_v\end{bmatrix}, \hspace*{0.5cm}\tilde{\mathscr{D}}_i=\tilde{D}_r\mathscr{D}_v.\nonumber
\end{align}
Let $\tilde{\mathscr{P}}_{12}$ solves the following Sylvester equation
\begin{align}
\tilde{A}_r\tilde{\mathscr{P}}_{12}+\tilde{\mathscr{P}}_{12}\mathscr{A}_v^T+\tilde{B}_r(\mathscr{C}_v\mathscr{P}_v+\mathscr{D}_v\mathscr{B}_v^T)=0.\nonumber
\end{align} Now define $\tilde{B}_{1}$ and $\tilde{B}_{2}$ as the following
\begin{align}
\tilde{B}_{1}=\begin{bmatrix}\tilde{B}_r&\tilde{\mathscr{P}}_{12}\mathscr{C}_v^T&\tilde{B}_r\mathscr{D}_v\end{bmatrix}\textnormal{ and }\tilde{B}_2=\begin{bmatrix}\tilde{\mathscr{P}}_{12}\mathscr{C}_v^T&\tilde{B}_r&\tilde{B}_r\mathscr{D}_v\end{bmatrix}.\nonumber
\end{align}
Then the frequency-weighted controllability Gramian $\tilde{P}_e$ of the pair $(\tilde{A}_r,\tilde{B}_r)$ solves the following Lyapunov equation
\begin{align}
\tilde{A}_r\tilde{P}_{e}+\tilde{P}_{e}\tilde{A}_r^T+\tilde{B}_1\tilde{B}_2^T=0.\nonumber
\end{align}
Now define $\mathscr{B}_{\mathscr{F}}$ and $\tilde{\mathscr{B}_{\mathscr{F}}}$ as
\begin{align}
\mathscr{B}_{\mathscr{F}}=\begin{bmatrix}\mathscr{P}_{12}\mathscr{C}_v^T+B\mathscr{D}_v\mathscr{D}_v^T\\\mathscr{P}_v\mathscr{C}_v^T+\mathscr{B}_v\mathscr{D}_v^T\end{bmatrix}\textnormal{ and }\tilde{\mathscr{B}_{\mathscr{F}}}=\begin{bmatrix}\tilde{\mathscr{P}}_{12}\mathscr{C}_v^T+\tilde{B}_r\mathscr{D}_v\mathscr{D}_v^T\\\mathscr{P}_v\mathscr{C}_v^T+\mathscr{B}_v\mathscr{D}_v^T\end{bmatrix}.\label{34F}
\end{align}
Let $\hat{X}$ and $\hat{Y}$ solve the following matrix equations
\begin{align}
\mathscr{A}_i\hat{X}&+\hat{X}\tilde{A}_r^T+\mathscr{B}_{\mathscr{F}}\tilde{B}_r^T=0,\label{39C}\\
\mathscr{A}_i^T\hat{Y}&+\hat{Y}\tilde{A}_r-\begin{bmatrix}C^T\\((D-\tilde{D}_r)\mathscr{C}_v)^T\end{bmatrix}\tilde{C}_r+\begin{bmatrix}0\\\mathscr{C}_v^T\end{bmatrix}\tilde{B}_r^T\tilde{Q}=0.\nonumber
\end{align}
Then $\tilde{G}_r(s)$ is a local-optimum for $||\big(G(s)-\tilde{G}_r(s)\big)\mathscr{V}(s)||^2_{\mathcal{H}_{2}}$ if it satisfies the following Halevi conditions \citep{halevi1992frequency}
\begin{align}
\mathscr{C}_i\hat{X}-\tilde{C}_r\tilde{P}_e-\tilde{D}_r\begin{bmatrix}0&\mathscr{C}_v\end{bmatrix}\hat{X}&=0\label{42C},\\
\hat{Y}^T\mathscr{B}_{\mathscr{F}}+\tilde{Q}\Big(\tilde{B}_r\mathscr{D}_v\mathscr{D}_v^T+\hat{X}\begin{bmatrix}0\\\mathscr{C}_v^T\end{bmatrix}\Big)&=0,\\
\hat{Y}^T\hat{X}+\tilde{Q}\tilde{P}_e&=0,\label{eq:8}\\
\tilde{C}_r\hat{X}^T\begin{bmatrix}0\\\mathscr{C}_v^T\end{bmatrix}M-C\mathscr{P}_{12}\mathscr{C}_v^TM-(D-\tilde{D}_r)\mathscr{C}_v\mathscr{P}_v\mathscr{C}_v^TM&=0,\label{44C}
\end{align} where $M$ is a basis for the null space of $\mathscr{D}_v^T$.

Let us define $\mathscr{F}[G(s)]$ and $\mathscr{F}[\tilde{G}_r(s)]$ as the following
\begin{align}
\mathscr{F}[G(s)]=\mathscr{C}_i(sI-\mathscr{A}_i)^{-1}\mathscr{B}_{\mathscr{F}}\textnormal{ and }\mathscr{F}[\tilde{G}_r(s)]=\tilde{\mathscr{C}}_{i}(sI-\tilde{\mathscr{A}}_{i})^{-1}\tilde{\mathscr{B}_{\mathscr{F}}}.\nonumber
\end{align}
Suppose $\tilde{G}_r(s)$ has only simple poles, and it can be represented in pole-residue form as
\begin{align}
\tilde{G}_r(s)=\sum_{i=1}^{r}\frac{c_ib_i^T}{s-\lambda_i}+\tilde{D}_r.\nonumber
\end{align}
Then the optimality conditions (\ref{42C})-(\ref{44C}) are equivalent to the following bi-tangential Hermite interpolation conditions, and a condition on $\tilde{D}_r$, i.e.,
\begin{align}
\mathscr{F}[G(-\lambda_i)]b_i&=\mathscr{F}[\tilde{G}_r(-\lambda_i)]b_i,\label{35C}\\
c_i^T\mathscr{F}[G(-\lambda_i)]&=c_i^T\mathscr{F}[\tilde{G}_r(-\lambda_i)],\\
c_i^T\mathscr{F}^\prime[G(-\lambda_i)]b_i&=c_i^T\mathscr{F}^\prime[\tilde{G}_r(-\lambda_i)]b_i,\label{37C}\\
\tilde{D}_r=C(\mathscr{P}_{12}-\tilde{V}_r\tilde{\mathscr{P}}_{12})&\mathscr{C}_v^TM(M^T\mathscr{C}_v\mathscr{P}_v\mathscr{C}_v^TM)^{-1}M^T.\label{38C}
\end{align}

The reduction subspaces $\tilde{V}_r$ and $\tilde{W}_r$ in NOWI are computed as
\begin{align}
Ran\begin{bmatrix}\tilde{V}^{(a)}\\\tilde{V}^{(b)}\end{bmatrix}&=\underset {i=1,\cdots,r}{span}\{(-\lambda_iI-\mathscr{A}_i)^{-1}\mathscr{B}_{\mathscr{F}} b_i\},\nonumber\\
Ran\begin{bmatrix}\tilde{W}^{(a)}\\\tilde{W}^{(b)}\end{bmatrix}&=\underset {i=1,\cdots,r}{span}\{(-\lambda_iI-\mathscr{A}_i^T)^{-1}\mathscr{C}_{i}^T c_i\},\nonumber\\
Ran(\tilde{V}_r)&\supset Ran(\tilde{V}^{(a)}),\textnormal{ and } Ran(\tilde{W}_r)\supset Ran(\tilde{W}^{(a)})\nonumber
\end{align} such that $\tilde{W}_r^T\tilde{V}_r=I$. $Ran(\cdot)$ represents the range of the matrix, $\underset {i=1,\cdots,r}{span}\{\cdot\}$ represents the span of the set of $r$ vectors, and $\supset$ represents the superset.

If $\mathscr{P}_{12}=\tilde{V}_r\tilde{\mathscr{P}}_{12}$, the interpolation conditions (\ref{35C})-(\ref{37C}) are enforced exactly; however, this is not always the case. Thus $\mathscr{P}_{12}-\tilde{V}_r\tilde{\mathscr{P}}_{12}$ quantifies the deviation in the interpolation conditions. The poles $\lambda_i$, right residues $b_i$, and left residues $c_i$ of $\tilde{G}_r(s)$ are not known \textit{a priori}. Therefore, an iterative algorithm similar to IRKA is proposed in \citep{breiten2015near}, which starts with a random set of interpolation points and tangential directions, and after each iteration, the interpolation points are updated as the mirror images of the poles of $\tilde{G}_r(s)$, and the tangential directions are updated as the respective right and left residues associated with these poles. At convergence, the ROM that nearly satisfies (\ref{35C})-(\ref{37C}) is obtained, and thus, NOWI generates a nearly optimal ROM.
\section{Main Work}
There is an important difference between the first-order optimality conditions for $||G(s)-\tilde{G}_r(s)||_{\mathcal{H}_2}$ and $||\mathscr{W}(s)\big(G(s)-\tilde{G}_r(s)\big)\mathscr{V}(s)||_{\mathcal{H}_{2}}$ problems. Since $\mathcal{H}_2$-norm of an improper transfer function becomes unbounded, $\tilde{D}_r$ is chosen to be equal to $D$ in the standard $\mathcal{H}_2$-MOR problem, and thus it is not involved in the optimality conditions. However, if $\tilde{D}_r\neq D$, $||\mathscr{W}(s)\big(G(s)-\tilde{G}_r(s)\big)\mathscr{V}(s)||_{\mathcal{H}_{2}}$ does not become unbounded as long as the transfer function $\mathscr{W}(s)\big(G(s)-\tilde{G}_r(s)\big)\mathscr{V}(s)$ is strictly proper. Therefore, the first-order optimality conditions for the frequency-weighted $\mathcal{H}_2$-MOR problem does involve a condition on the selection of $\tilde{D}_r$ (see the equations (\ref{44C}) and (\ref{38C})).

In this section, we focus on the frequency-weighted $\mathcal{H}_2$-pseudo-optimal MOR problem. Let us divide this problem into three cases before defining the notion of pseudo-optimality:
\begin{align}
&\textnormal{(i) } \underset{\substack{\tilde{G}_r(s)\\\mathscr{W}(s)=I}}{\text{min}}||\mathscr{W}(s)\big(G(s)-\tilde{G}_r(s)\big)\mathscr{V}(s)||^2_{\mathcal{H}_{2}}.\nonumber\\
&\textnormal{(ii) } \underset{\substack{\tilde{G}_r(s)\\\mathscr{V}(s)=I}}{\text{min}}||\mathscr{W}(s)\big(G(s)-\tilde{G}_r(s)\big)\mathscr{V}(s)||^2_{\mathcal{H}_{2}}.\nonumber\\
&\textnormal{(iii) }  \underset{\substack{\tilde{G}_r(s)\\\mathscr{V}(s)\neq I,\mathscr{W}(s)\neq I}}{\text{min}}||\mathscr{W}(s)\big(G(s)-\tilde{G}_r(s)\big)\mathscr{V}(s)||^2_{\mathcal{H}_{2}}.\nonumber
\end{align}

By pseudo-optimality, we mean satisfying the first-order optimality conditions for a subset of the state-space matrices of the ROM, i.e., we restrict ourselves to achieving the first-order optimality conditions associated with $\tilde{B}_r$ or $\tilde{C}_r$, i.e.,
\begin{align}
\frac{\partial}{\partial\tilde{C}_r}&\Big(||\big(G(s)-\tilde{G}_r(s)\big)\mathscr{V}(s)||^2_{\mathcal{H}_{2}}\Big)=0,\label{49C}\\
\frac{\partial}{\partial\tilde{B}_r}&\Big(||\mathscr{W}(s)\big(G(s)-\tilde{G}_r(s)\big)||^2_{\mathcal{H}_{2}}\Big)=0.\label{50C}
\end{align}
We name our algorithms as pseudo-optimal weighted-interpolation (POWI) algorithms. We propose two iteration-free weighted interpolation algorithms for the cases (i) and (ii), which constructs a ROM\\ ($\tilde{A}_r,\tilde{B}_r,\tilde{C}_r,\tilde{D}_r$) that satisfies (\ref{49C}) and (\ref{50C}), respectively. We refer these as Input-sided POWI (I-POWI) and Output-sided POWI (O-POWI) for differentiation. The stability of the ROM is guaranteed in I-POWI and O-POWI. For the cases (iii), we propose an IRKA-like iterative algorithm that tends to achieve the optimality condition (\ref{49C}) and (\ref{50C}). We refer to this algorithm as Double-sided POWI (D-POWI). Note that D-POWI closely resembles the heuristic algorithm presented in \citep{zulfiqar2018weighted}, which is based on analogy and experimental results. D-POWI, however, is based on the first-order optimality conditions (\ref{49C}) and (\ref{50C}). I-POWI satisfies the optimality condition that NOWI fails to satisfy, whereas D-POWI is the double-sided extension of NOWI. Moreover, POWI algorithms are computationally efficient than existing techniques.
\subsection{I-POWI}
In this subsection, we consider the FWMOR problem
\begin{align}
\underset{\substack{\tilde{G}_r(s)\\\mathscr{W}(s)=I}}{\text{min}}||\mathscr{W}(s)\big(G(s)-\tilde{G}_r(s)\big)\mathscr{V}(s)||^2_{\mathcal{H}_{2}}.\nonumber
\end{align}
If $\tilde{D}_r=D$, the state-space realization of $\big(G(s)-\tilde{G}_r(s)\big)\mathscr{V}(s)$ can be written as
\begin{align}
\mathscr{E}_i(s)=\big(G(s)-\tilde{G}_r(s)\big)\mathscr{V}(s)=\mathscr{C}_{e,i}(sI-\mathscr{A}_{e,i})^{-1}\mathscr{B}_{e,i}\nonumber
\end{align} where
\begin{align}
\mathscr{A}_{e,i}&=\begin{bmatrix}A&0&B\mathscr{C}_{v}\\0&\tilde{A}_r&\tilde{B}_r\mathscr{C}_{v}\\0&0&\mathscr{A}_{v}\end{bmatrix},
&&\mathscr{B}_{e,i}=\begin{bmatrix}B\mathscr{D}_v\\\tilde{B}_r\mathscr{D}_v\\\mathscr{B}_v\end{bmatrix},&&\mathscr{C}_{e,i}=\begin{bmatrix}C&-\tilde{C}_r&0\end{bmatrix}.\nonumber
\end{align} The controllability Gramian of the pair $(\mathscr{A}_{e,i},\mathscr{B}_{e,i})$ satisfies the following Lyapunov equation
\begin{align}
\mathscr{A}_{e,i}\mathscr{P}_{e,i}+\mathscr{P}_{e,i}\mathscr{A}_{e,i}^T+\mathscr{B}_{e,i}\mathscr{B}_{e,i}^T=0.\nonumber
\end{align} $\mathscr{P}_{e,i}$ can be partitioned according to $\mathscr{A}_{e,i}$ as
\begin{align}
\mathscr{P}_{e,i}=\begin{bmatrix}P_e&\hat{\mathscr{P}}_{12}&\mathscr{P}_{12}\\\hat{\mathscr{P}}_{12}^T&\tilde{P}_e&\tilde{\mathscr{P}}_{12}\\\mathscr{P}_{12}^T&\tilde{\mathscr{P}}_{12}^T&\mathscr{P}_{v}\end{bmatrix}\nonumber
\end{align} where $\hat{\mathscr{P}}_{12}$ solves the following Sylvester equation
\begin{align}
A\hat{\mathscr{P}}_{12}+\hat{\mathscr{P}}_{12}\tilde{A}_{r}^T+B_{1}\tilde{B}_{2}^T=0.\nonumber
\end{align}
Now
\begin{align}
||\mathscr{E}_i(s)&||_{\mathcal{H}_2}=\sqrt{tr(\mathscr{C}_{e,i}\mathscr{P}_{e,i}\mathscr{C}_{e,i}^T)}\nonumber\\
&=\sqrt{tr(CP_eC^T)-2tr(C\hat{\mathscr{P}}_{12}\tilde{C}_r^T)+tr(\tilde{C}_r\tilde{P}_e\tilde{C}_r^T)}\nonumber
\end{align} where $tr(\cdot)$ represents the trace of the matrix.
The cost function $J_i(\tilde{A}_r,\tilde{B}_r,\tilde{C}_r)$ for minimizing $||\mathscr{E}_i(s)||^2_{\mathcal{H}_2}$ can be written as
\begin{align}
J_i=-2tr(C\hat{\mathscr{P}}_{12}\tilde{C}_r^T)+tr(\tilde{C}_r\tilde{P}_e\tilde{C}_r^T).\nonumber
\end{align}
The partial derivative of $J_i$ with respect to $\tilde{C}_r$ is given by
\begin{align}
\Delta J_{\tilde{C}_r}=2\tilde{C}_r\tilde{P}_e-2C\hat{\mathscr{P}}_{12}.\nonumber
\end{align}
Then the first-order optimality condition associated with $\tilde{C}_r$ becomes
\begin{align}
\tilde{C}_r\tilde{P}_e-C\hat{\mathscr{P}}_{12}=0.\label{51C}
\end{align}
It can readily be shown that (\ref{51C}) is equivalent to the second Halevi condition, i.e., (\ref{42C}) when $\tilde{D}_r=D$. One can notice by expanding the equation (\ref{39C}) that $\hat{X}=\begin{bmatrix}\hat{\mathscr{P}}_{12}\\\tilde{\mathscr{P}}_{12}^T\end{bmatrix}$. Thus the equation (\ref{42C}) becomes
\begin{align}
\begin{bmatrix}C&D\mathscr{C}_v\end{bmatrix}\begin{bmatrix}\hat{\mathscr{P}}_{12}\\\tilde{\mathscr{P}}_{12}^T\end{bmatrix}-\tilde{C}_r\tilde{P}_e-D\begin{bmatrix}0&\mathscr{C}_v\end{bmatrix}\begin{bmatrix}\hat{\mathscr{P}}_{12}\\\tilde{\mathscr{P}}_{12}^T\end{bmatrix}=0\nonumber\\
C\hat{\mathscr{P}}_{12}-\tilde{C}_r\tilde{P}_e=0.\nonumber
\end{align}
Moreover, since (\ref{42C}) is equivalent to (\ref{35C}), the condition (\ref{51C}) is also equivalent to (\ref{35C}) when $\tilde{D}_r=D$. We now present an algorithm which minimizes $||\mathscr{E}_i(s)||^2_{\mathcal{H}_2}$ with respect to $\tilde{C}_r$, i.e., the ROM satisfies (\ref{51C}). Let $\{\sigma_1,\cdots,\sigma_r\}$ be the interpolation points with the respective right tangential directions $\{\hat{r}_1,\cdots,\hat{r}_r\}$ where $\{\sigma_1,\cdots,\sigma_r\}$ lie in the right-half of the $s$-plane. Then the input rational Krylov subspace $\tilde{V}_a=\begin{bmatrix}\tilde{V}_r^T&\tilde{V}_b^T\end{bmatrix}^T$ can be obtained as
\begin{align}
Ran\Bigg(\begin{bmatrix}\tilde{V}_r\\\tilde{V}_b\end{bmatrix}\Bigg)&=\underset {i=1,\cdots,r}{span}\{(\sigma_iI-\mathscr{A}_i)^{-1}\mathscr{B}_{\mathscr{F}} \hat{r}_i\}.\label{52E}
\end{align}
Choose any $\tilde{W}_a$ such that $\tilde{W}_a^T\tilde{V}_a=I$; for instance, $\tilde{W}_a=\tilde{V}_a(\tilde{V}_a^T\tilde{V}_a)^{-1}$ and define the following matrices
\begin{align}
\hat{A}&=\tilde{W}_a^T\mathscr{A}_i\tilde{V}_a,\hspace*{0.3cm}\hat{B}=\tilde{W}_a^T\mathscr{B}_{\mathscr{F}},\hspace*{0.3cm}B_\bot=\mathscr{B}_{\mathscr{F}}-\tilde{V}_a\hat{B},\label{53E}\\
L_i&=(B_\bot^TB_\bot)^{-1}B_\bot^T\Big(\mathscr{A}_i\tilde{V}_a-\tilde{V}_a\hat{A}\Big),\hspace*{0.3cm}S=\hat{A}-\hat{B}L_i.\label{54E}
\end{align}
Owing to the connection between rational Krylov subspace and Sylvester equation, $\tilde{V}_a$ solves the following Sylvester equation
\begin{align}
\mathscr{A}_i\tilde{V}_a-\tilde{V}_aS-\mathscr{B}_{\mathscr{F}}L_i=0\label{52C}
\end{align} where $\{\sigma_1,\cdots,\sigma_r\}$ are the eigenvalues of $S$; see \citep{panzer2014model,wolf2014h} for details.
The equation (\ref{52C}) can be expanded according to $\mathscr{A}_i$ as
\begin{align}
A\tilde{V}_r-\tilde{V}_rS+B\mathscr{C}_v\tilde{V}_b-(\mathscr{P}_{12}\mathscr{C}_v^T+B\mathscr{D}_v\mathscr{D}_v^T)L_i=0,\label{53C}\\
\mathscr{A}_v\tilde{V}_b-\tilde{V}_bS-(\mathscr{P}_v\mathscr{C}_v^T+\mathscr{B}_v\mathscr{D}_v^T)L_i=0.
\end{align}
Let $\mathscr{P}_s$ be the frequency-weighted controllability Gramian of the pair $(-S^T,-L_i^T)$. Then clearly $\mathscr{P}_s$ solves the following Lyapunov equation
\begin{align}
-S^T\mathscr{P}_{s}-\mathscr{P}_{s}S+L_{i,1}L_{i,2}^T=0\label{60E}
\end{align} where
\begin{align}
L_{i,1}=\begin{bmatrix}-L_i^T&\tilde{V}_b^T\mathscr{C}_v^T&-L_i^T\mathscr{D}_v\end{bmatrix},\label{61F}\\
L_{i,2}=\begin{bmatrix}\tilde{V}_b^T\mathscr{C}_v^T&-L_i^T&-L_i^T\mathscr{D}_v\end{bmatrix}.\label{62F}
\end{align} Note that the computational cost of $\mathscr{P}_{s}$ is dominated by the small-scale matrix $S$, and hence, it can be computed efficiently. Moreover, the reuse of $\tilde{V}_b$ in (\ref{61F}) and (\ref{62F}) for the computation of $\mathscr{P}_{s}$ also save computational effort.

Then the ROM in I-POWI is constructed as
\begin{align}
\tilde{A}_r&=-\mathscr{P}_{s}^{-1}S^T\mathscr{P}_{s},&&\tilde{B}_r=-\mathscr{P}_{s}^{-1}L_i^T\nonumber\\
\tilde{C}_r&=C\tilde{V}_r,&&\tilde{D}_r=D.\label{58C}
\end{align}
\textbf{Theorem 1:} If $(\tilde{A}_r,\tilde{B}_r,\tilde{C}_r,\tilde{D}_r)$ is computed as in the equation (\ref{58C}), the following statements are true:\\
(i) $\mathscr{P}_{s}^{-1}$ is the frequency-weighted controllability Gramian of the pair $(\tilde{A}_r,\tilde{B}_r)$.\\
(ii) $\tilde{\mathscr{P}}_{12}=\mathscr{P}_{s}^{-1}\tilde{V}_b^T$.\\
(iii) $\tilde{G}_r(s)$ satisfies the condition (\ref{51C}).\\
(iv) The poles $\lambda_i$ of $\tilde{G}_r(s)$ are at the mirror images of the interpolation points.\\
\textbf{Proof:} (i) By applying a state transformation on (\ref{58C}) using $\mathscr{P}_{s}^{-1}$ as the transformation matrix, we get
\begin{align}
\tilde{A}_{t,r}&=-S^T, && \tilde{B}_{t,r}=-L_i^T,\nonumber\\
\tilde{C}_{t,r}&=C\tilde{V}_r\mathscr{P}_{s}^{-1},&& \tilde{D}_{t,r}=D.
\end{align}
Note that the frequency-weighted controllability Gramian of the pair $(\tilde{A}_{t,r},\tilde{B}_{t,r})$ is equal to $\mathscr{P}_{s}$. Now the frequency-weighted controllability Gramian $\tilde{P}_{e}$ of the pair $(\tilde{A}_r,\tilde{B}_r)$ can be recovered from that of the pair $(\tilde{A}_{t,r},\tilde{B}_{t,r})$ using the similarity transformation matrix $\mathscr{P}_{s}^{-1}$ as $\tilde{P}_{e}=\mathscr{P}_{s}^{-1}\mathscr{P}_{s}\mathscr{P}_{s}^{-1}=\mathscr{P}_{s}^{-1}$.\\
(ii) Consider the following Sylvester equation
\begin{align}
&\tilde{A}_r\mathscr{P}_s^{-1}\tilde{V}_b^T+\mathscr{P}_s^{-1}\tilde{V}_b^T\mathscr{A}_v^T+\tilde{B}_r(\mathscr{C}_v\mathscr{P}_v+\mathscr{D}_v\mathscr{B}_v^T)\nonumber\\
=&-\mathscr{P}_s^{-1}S^T\tilde{V}_b^T+\mathscr{P}_s^{-1}\tilde{V}_b^T\mathscr{A}_v^T-\mathscr{P}_s^{-1}L_i^T(\mathscr{C}_v\mathscr{P}_v+\mathscr{D}_v\mathscr{B}_v^T)\nonumber\\
=&\mathscr{P}_s^{-1}\big(\tilde{V}_b^T\mathscr{A}_v^T-S^T\tilde{V}_b^T-L_i^T(\mathscr{C}_v\mathscr{P}_v+\mathscr{D}_v\mathscr{B}_v^T)\big)=0.\nonumber
\end{align}
Due to uniqueness $\mathscr{P}_s^{-1}\tilde{V}_b^T=\tilde{\mathscr{P}}_{12}$.\\
(iii) Consider the following Sylvester equation
\begin{align}
&A\tilde{V}_r\tilde{P}_e+\tilde{V}_r\tilde{P}_e\tilde{A}_{r}^T+B\mathscr{C}_v\tilde{\mathscr{P}}_{12}^T+\mathscr{P}_{12}\mathscr{C}_v^T\tilde{B}_r^T+B\mathscr{D}_v\mathscr{D}_v^T\tilde{B}_r\nonumber\\
=&[\tilde{V}_rS-B\mathscr{C}_v\tilde{V}_b+\mathscr{P}_{12}\mathscr{C}_v^TL_i+B\mathscr{D}_v\mathscr{D}_v^TL_i]\tilde{P}_e-\tilde{V}_rS\tilde{P}_e+B\mathscr{C}_v\tilde{\mathscr{P}}_{12}^T-\mathscr{P}_{12}\mathscr{C}_v^TL_i\tilde{P}_e-B\mathscr{D}_v\mathscr{D}_v^TL_i\tilde{P}_e\nonumber\\
=&[\tilde{V}_rS-B\mathscr{C}_v\tilde{\mathscr{P}}_{12}^T\tilde{P}_e^{-1}+\mathscr{P}_{12}\mathscr{C}_v^TL_i+B\mathscr{D}_v\mathscr{D}_v^TL_i]\tilde{P}_e-\tilde{V}_rS\tilde{P}_e+B\mathscr{C}_v\tilde{\mathscr{P}}_{12}^T-\mathscr{P}_{12}\mathscr{C}_v^TL_i\tilde{P}_e-B\mathscr{D}_v\mathscr{D}_v^TL_i\tilde{P}_e\nonumber\\
=&0.\nonumber\end{align}
Due to uniqueness, $\tilde{V}_r\tilde{P}_e=\hat{\mathscr{P}}_{12}$, and thus $\tilde{C}_r\tilde{P}_e=C\hat{\mathscr{P}}_{12}$.\\
(iv) Since $\tilde{A}_r=-\mathscr{P}_{s}^{-1}S^T\mathscr{P}_{s}$, $\lambda_i(\tilde{A}_r)=\lambda_i(-S^T)$ where $\lambda_i(\cdot)$ represents the eigenvalues of the matrix. Thus, $\tilde{G}_r(s)$ has poles at $-\sigma_i$.\\

Now recall
\begin{align}
||\mathscr{E}_i(s)||^2_{\mathcal{H}_2}=tr(CP_eC^T)-2tr(C\hat{\mathscr{P}}_{12}\tilde{C}_r^T)+tr(\tilde{C}_r\tilde{P}_e\tilde{C}_r^T).\nonumber
\end{align} Since the ROM generated by I-POWI ensures that $\tilde{C}_r\tilde{P}_e=C\hat{\mathscr{P}}_{12}$,
\begin{align}
||\mathscr{E}_i(s)||^2_{\mathcal{H}_2}&=tr(CP_eC^T)-tr(\tilde{C}_r\tilde{P}_e\tilde{C}_r^T)\nonumber\\
&=tr\big(C(P_e-\tilde{V}_r\tilde{P}_e\tilde{V}_r^T)C^T\big)\nonumber\\
&=tr\big(C(P_e-\tilde{V}_r\mathscr{P}_s^{-1}\tilde{V}_r^T)C^T\big).\nonumber
\end{align} Thus I-POWI implicitly generates an approximation of $P_e$, i.e., $\hat{P}_e=\tilde{V}_r\mathscr{P}_s^{-1}\tilde{V}_r^T$. $P_e$ can be replaced with $\hat{P}_e$ in FWBT to construct a ROM, and the solution of large-scale Lyapunov equation (\ref{17C}) can be avoided to save computational effort. A pseudo-code for I-POWI is given in Algorithm 1.\\
\\\textbf{Algorithm 1: I-POWI}\\
\textbf{\textit{Input:}} Original system : $(A,B,C,D)$; input weight $(\mathscr{A}_v,\mathscr{B}_v,\mathscr{C}_v,\mathscr{D}_v)$; interpolation points $(\sigma_1,\cdots,\sigma_r)$; right tangential directions $(\hat{r}_1,\cdots,\hat{r}_r)$.\\
\textbf{\textit{Output:}} ROM: $(\tilde{A}_r,\tilde{B}_r,\tilde{C}_r,\tilde{D}_r)$.\\
(i) Compute $\mathscr{P}_v$ and $\mathscr{P}_{12}$ from the equations (\ref{7}) and (\ref{18C}), respectively.\\
(ii) Define $\mathscr{A}_i$ and $\mathscr{B}_{\mathscr{F}}$ according to the equations (\ref{13F}) and (\ref{34F}), respectively.\\
(iii) Compute $\tilde{V}_a=\begin{bmatrix}\tilde{V}_r^T&\tilde{V}_b^T\end{bmatrix}^T$ from the equation (\ref{52E}).\\
(iv) Set $\tilde{W}_a=\tilde{V}_a(\tilde{V}_a^T\tilde{V}_a)^{-1}$, and compute $S$ and $L_i$ from the equations (\ref{53E}) and (\ref{54E}), respectively.\\
(v) Define $L_{i,1}$ and $L_{i,2}$ according to the equations (\ref{61F}) and (\ref{62F}), respectively.\\
(vi) Compute $\mathscr{P}_s$ from the equation (\ref{60E}).\\
(vii) Compute the ROM from the equation (\ref{58C}).\\
\\\textbf{Remark 1:} When $\mathscr{V}(s)=I$, I-POWI becomes equivalent to PORK \citep{wolf2014h}, and when $\mathscr{V}(s)$ is an ideal (infinite-order) bandpass filter, it becomes equivalent to frequency-limited PORK (FLPORK) \citep{zulfiqar2020frequency}.
\subsection{O-POWI}
In this subsection, we consider the FWMOR problem
\begin{align}
\underset{\substack{\tilde{G}_r(s)\\\mathscr{V}(s)=I}}{\text{min}}||\mathscr{W}(s)\big(G(s)-\tilde{G}_r(s)\big)\mathscr{V}(s)||^2_{\mathcal{H}_{2}}.\nonumber
\end{align}
The state-space realization of the output augmented system $\mathscr{W}(s)\tilde{G}_r(s)$ can be written as the following
\begin{align}
\tilde{\mathscr{H}}_o(s)=\mathscr{W}(s)\tilde{G}_r(s)=\tilde{\mathscr{C}}_o(sI-\tilde{\mathscr{A}}_o)^{-1}\tilde{\mathscr{B}}_o+\tilde{\mathscr{D}}_o\nonumber
\end{align} where
\begin{align}
\tilde{\mathscr{A}}_o=&\begin{bmatrix}\tilde{A}_r&0\\\mathscr{B}_w\tilde{C}_r&\mathscr{A}_w\end{bmatrix}, \hspace*{0.5cm}\tilde{\mathscr{B}}_o=\begin{bmatrix}\tilde{B}_r\\\mathscr{B}_wD\end{bmatrix},\nonumber\\
\tilde{\mathscr{C}}_o=&\begin{bmatrix}\mathscr{D}_w\tilde{C}_r &\mathscr{C}_w\end{bmatrix}, \hspace*{0.5cm}\tilde{\mathscr{D}}_o=\mathscr{D}_wD.\nonumber
\end{align}
The observability Gramian $\tilde{\mathscr{Q}}_o$ of the pair $(\tilde{\mathscr{A}}_o,\tilde{\mathscr{C}}_o)$ is the solution of the following Lyapunov equation
\begin{align}
\tilde{\mathscr{A}}_o^T\tilde{\mathscr{Q}}_o+\tilde{\mathscr{Q}}_o\tilde{\mathscr{A}}_o+\tilde{\mathscr{C}}_o^T\tilde{\mathscr{C}}_o=0.\label{62C}
\end{align}
$\tilde{\mathscr{Q}}_o$ can be partitioned according to $\tilde{\mathscr{A}}_o$ as
\begin{align}
\tilde{\mathscr{Q}}_o=\begin{bmatrix}\tilde{Q}_e&\tilde{\mathscr{Q}}_{12}\\\tilde{\mathscr{Q}}_{12}^T&\mathscr{Q}_w\end{bmatrix}.\label{63C}
\end{align}
Let us define $\tilde{C}_{1}^T$ and $\tilde{C}_{2}^T$ as the following
\begin{align}
\tilde{C}_{1}^T=\begin{bmatrix}\tilde{C}_r^T&\tilde{\mathscr{Q}}_{12}\mathscr{B}_w&\tilde{C}_r^T\mathscr{D}_w^T\end{bmatrix}\textnormal{ and }\tilde{C}_{2}^T=\begin{bmatrix}\tilde{\mathscr{Q}}_{12}\mathscr{B}_w&\tilde{C}_r^T&\tilde{C}_r^T\mathscr{D}_w^T\end{bmatrix}.\nonumber
\end{align}
Then it can be noted by expanding the equation (\ref{62C}) according to (\ref{63C}) that $\tilde{\mathscr{Q}}_{12}$ and $\tilde{Q}_e$ solve the following matrix equations
\begin{align}
\tilde{A}_r^T\tilde{\mathscr{Q}}_{12}+\tilde{\mathscr{Q}_{12}}\mathscr{A}_w+\tilde{C}_r^T(\mathscr{B}_w^T\mathscr{Q}_w+\mathscr{D}_w^T\mathscr{C}_w)=0\\
\tilde{A}_r^T\tilde{Q}_e+\tilde{Q}_e\tilde{A}_r+\tilde{C}_{1}^T\tilde{C}_{2}=0.
\end{align}
$\tilde{Q}_e$ is the frequency-weighted observability Gramian of the pair $(\tilde{A}_r,\tilde{C}_r)$.

If $\tilde{D}_r=D$, the state-space realization of the output augmented system $\mathscr{W}(s)\big(G(s)-\tilde{G}_r(s)\big)$ can be written as the following
\begin{align}
\mathscr{E}_o(s)=\mathscr{W}(s)\big(G(s)-\tilde{G}_r(s)\big)=\mathscr{C}_{e,o}(sI-\mathscr{A}_{e,o})^{-1}\mathscr{B}_{e,o}\nonumber
\end{align} where
\begin{align}
\mathscr{A}_{e,o}&=\begin{bmatrix}A&0&0\\0&\tilde{A}_r&0\\\mathscr{B}_wC&-\mathscr{B}_w\tilde{C}_r&\mathscr{A}_w\end{bmatrix}, \hspace*{0.5cm}\mathscr{B}_{e,o}=\begin{bmatrix}B\\\tilde{B}_r\\0\end{bmatrix},\hspace*{0.5cm}\mathscr{C}_{e,o}=\begin{bmatrix}\mathscr{D}_wC&-\mathscr{D}_w\tilde{C}_r&\mathscr{C}_w\end{bmatrix}.\nonumber
\end{align}
The observability Gramian $\mathscr{Q}_{e,o}$ of the pair $(\mathscr{A}_{e,o},\mathscr{C}_{e,o})$ satisfies the following Lyapunov equation
\begin{align}
\mathscr{A}_{e,o}^T\mathscr{Q}_{e,o}+\mathscr{Q}_{e,o}\mathscr{A}_{e,o}+\mathscr{C}_{e,o}^T\mathscr{C}_{e,o}=0.\nonumber
\end{align}
$\mathscr{Q}_{e,o}$ can be partitioned according to $\mathscr{A}_{e,o}$ as
\begin{align}
\mathscr{Q}_{e,o}=\begin{bmatrix}Q_e&-\hat{\mathscr{Q}_{12}}&\mathscr{Q}_{12}\\-\hat{\mathscr{Q}}_{12}^T&\tilde{Q}_e&\tilde{\mathscr{Q}}_{12}\\\mathscr{Q}_{12}^T&\tilde{\mathscr{Q}}_{12}^T&\mathscr{Q}_w\end{bmatrix}\nonumber
\end{align} where $\hat{\mathscr{Q}_{12}}$ solves the following Sylvester equation
\begin{align}
A^T\hat{\mathscr{Q}_{12}}+\hat{\mathscr{Q}_{12}}\tilde{A}_r+C_1^T\tilde{C}_2=0.\nonumber
\end{align}
Now
\begin{align}
||\mathscr{E}_o(s)&||_{\mathcal{H}_2}=\sqrt{tr(\mathscr{B}_{e,o}^T\mathscr{Q}_{e,o}\mathscr{B}_{e,o})}\nonumber\\
&=\sqrt{tr(B^TQ_eB)-2tr(B^T\hat{\mathscr{Q}}_{12}\tilde{B}_r)+tr(\tilde{B}_r^T\tilde{Q}_e\tilde{B}_r)}.\nonumber
\end{align}
The cost function $J_o(\tilde{A}_r,\tilde{B}_r,\tilde{C}_r)$ for minimizing $||\mathscr{E}_o(s)||^2_{\mathcal{H}_2}$ can be written as
\begin{align}
J_o=-2tr(B^T\hat{\mathscr{Q}}_{12}\tilde{B}_r)+tr(\tilde{B}_r^T\tilde{Q}_e\tilde{B}_r).\nonumber
\end{align}
The partial derivative of $J_o$ with respect to $\tilde{B}_r$ is given by
\begin{align}
\Delta J_{\tilde{B}_r}=2\tilde{B}_r^T\tilde{Q}_e-2B^T\hat{\mathscr{Q}}_{12}.\nonumber
\end{align}
Then the first-order optimality condition associated with $\tilde{B}_r$ becomes
\begin{align}
\tilde{Q}_e\tilde{B}_r-\hat{\mathscr{Q}}_{12}^TB=0.\label{68C}
\end{align}
We know that the controllability Gramian of the pair $(A^T,C^T)$ is equal to the observability Gramian of the pair $(A,C)$. By noting this duality, one can readily notice that (\ref{68C}) is a dual of (\ref{51C}). This is intuitive because $\mathscr{E}_o^T(s)=\big(G^T(s)-\tilde{G}_r^T(s)\big)\mathscr{W}^T(s)$. Now let us define the dual realizations of $\mathscr{F}[G(s)]$ and $\mathscr{F}[\tilde{G}_r(s)]$, and denote these as $\mathscr{G}[G(s)]$ and $\mathscr{G}[\tilde{G}_r(s)]$, respectively, i.e.,
\begin{align}
\mathscr{G}[G(s)]=\mathscr{C}_{\mathscr{G}}(sI-\mathscr{A}_o)^{-1}\mathscr{B}_o\textnormal{ and }\mathscr{G}[\tilde{G}_r(s)]=\tilde{\mathscr{C}}_{\mathscr{G}}(sI-\tilde{\mathscr{A}}_o)^{-1}\tilde{\mathscr{B}}_o\nonumber
\end{align} where
\begin{align}
\mathscr{C}_{\mathscr{G}}=\begin{bmatrix}\mathscr{B}_w^T\mathscr{Q}_{12}^T+\mathscr{D}_w^T\mathscr{D}_wC&\mathscr{B}_w^T\mathscr{Q}_w+\mathscr{D}_w^T\mathscr{C}_w\end{bmatrix}\textnormal{ and }\tilde{\mathscr{C}}_{\mathscr{G}}=\begin{bmatrix}\mathscr{B}_w^T\tilde{\mathscr{Q}}_{12}^T+\mathscr{D}_w^T\mathscr{D}_w\tilde{C}_r&\mathscr{B}_w^T\mathscr{Q}_w+\mathscr{D}_w^T\mathscr{C}_w\end{bmatrix}.\label{71F}
\end{align}
Due to duality, (\ref{68C}) becomes equivalent to achieving the interpolation condition \begin{align}
c_i^T\mathscr{G}[G(-\lambda_i)]=c_i^T\mathscr{G}[\tilde{G}_r(-\lambda_i)]\nonumber
\end{align}
when $\tilde{D}_r=D$. We now present an algorithm that minimizes $||\mathscr{E}_o(s)||^2_{\mathcal{H}_2}$ with respect to $\tilde{B}_r$, i.e., the ROM satisfies (\ref{68C}).
Let $\{\sigma_1,\cdots,\sigma_r\}$ be the interpolation points with the respective left tangential directions $\{\hat{l}_1,\cdots,\hat{l}_r\}$ where $\{\sigma_1,\cdots,\sigma_r\}$ lie in the right-half of the $s$-plane. Then the output rational Krylov subspace $\tilde{W}_a=\begin{bmatrix}\tilde{W}_r^T&\tilde{W}_b^T\end{bmatrix}^T$ can be obtained as
\begin{align}
Ran\Bigg(\begin{bmatrix}\tilde{W}_r\\\tilde{W}_b\end{bmatrix}\Bigg)&=\underset {i=1,\cdots,r}{span}\{(\sigma_iI-\mathscr{A}_o^T)^{-1}\mathscr{C}_{\mathscr{G}}^T \hat{l}_i\}.\label{75F}
\end{align}
Set $\tilde{V}_o=\tilde{W}_a$, $\tilde{W}_o=\tilde{V}_o(\tilde{V}_o^T\tilde{V}_o)^{-1}$, and define the following matrices
\begin{align}
\hat{A}&=\tilde{W}_o^T\mathscr{A}_o\tilde{V}_o,\hspace*{0.2cm}\hat{C}=\mathscr{C}_{\mathscr{G}}\tilde{V}_o, \hspace*{0.2cm}C_\bot=\mathscr{C}_{\mathscr{G}}-\hat{C}\tilde{W}_o^T,\label{76F}\\
L_o&=\big(\tilde{W}_o^TA-\hat{A}\tilde{W}_o^T\big)C_\bot^T(C_\bot C_\bot^T)^{-1},\hspace*{0.2cm}S=\hat{A}-L_o\hat{C}.\label{77F}
\end{align}
Owing to the connection between rational Krylov subspace and Sylvester equation, $\tilde{W}_a$ solves the following Sylvester equation
\begin{align}
-S\tilde{W}_a^T+\tilde{W}_a^T\mathscr{A}_o-L_o\mathscr{C}_{\mathscr{G}}=0\label{69C}
\end{align} where $\{\sigma_1,\cdots,\sigma_r\}$ are the eigenvalues of $S$; see \citep{panzer2014model} for details. The equation (\ref{69C}) can be expanded according to $\mathscr{A}_o$ as
\begin{align}
&-S\tilde{W}_r^T+\tilde{W}_r^TA+\tilde{W}_b^T\mathscr{B}_wC-L_o(\mathscr{B}_w^T\mathscr{Q}_{12}^T+\mathscr{D}_w^T\mathscr{D}_wC)=0\label{70C}\\
&-S\tilde{W}_b^T+\tilde{W}_b^T\mathscr{A}_w-L_o(\mathscr{B}_w^T\mathscr{Q}_w+\mathscr{D}_w^T\mathscr{C}_w)=0.
\end{align}

Let $\mathscr{Q}_s$ be the frequency-weighted observability Gramian of the pair $(-S^T,-L_o^T)$. Then it solves the following Lyapunov equation
\begin{align}
-S\mathscr{Q}_s-\mathscr{Q}_sS^T+L_{o,1}^TL_{o,2}=0\label{84F}
\end{align} where
\begin{align}
L_{o,1}^T=\begin{bmatrix}-L_o&\tilde{W}_b^T\mathscr{B}_w&-L_o\mathscr{D}_w^T\end{bmatrix},\label{85F}\\
L_{o,2}^T=\begin{bmatrix}\tilde{W}_b^T\mathscr{B}_w&-L_o&-L_o\mathscr{D}_w^T\end{bmatrix}.\label{86F}
\end{align} Note that the computational cost of $\mathscr{Q}_{s}$ is dominated by the small-scale matrix $S$, and hence, it can be computed efficiently. Moreover, the reuse of $\tilde{W}_b$ in (\ref{85F}) and (\ref{86F}) for the computation of $\mathscr{Q}_{s}$ also save computational effort.

Then the ROM in O-POWI is constructed as
\begin{align}
\tilde{A}_r&=-\mathscr{Q}_{s}S^T\mathscr{Q}_{s}^{-1},&&\tilde{B}_r=\tilde{W}_r^TB\nonumber\\
\tilde{C}_r&=-L_o^T\mathscr{Q}_{s}^{-1},&&\tilde{D}_r=D.\label{78D}
\end{align}
\textbf{Theorem 2:} If $(\tilde{A}_r,\tilde{B}_r,\tilde{C}_r,\tilde{D}_r)$ is computed as in the equation (\ref{78D}), the following statements are true:\\
(i) $\mathscr{Q}_{s}^{-1}$ is the frequency-weighted observability Gramian of the pair $(\tilde{A}_r,\tilde{C}_r)$.\\
(ii) $\tilde{\mathscr{Q}}_{12}=\mathscr{Q}_{s}^{-1}\tilde{W}_b^T$.\\
(iii) $\tilde{G}_r(s)$ satisfies the condition (\ref{68C}).\\
(iv) The poles $\lambda_i$ of $\tilde{G}_r(s)$ are at the mirror images of the interpolation points.\\
\textbf{Proof:} (i) By applying a state transformation on (\ref{78D}) using $\mathscr{Q}_{s}$ as the transformation matrix, we get
\begin{align}
\tilde{A}_{t,r}&=-S^T, && \tilde{B}_{t,r}=\mathscr{Q}_{s}^{-1}\tilde{W}_rB,\nonumber\\
\tilde{C}_{t,r}&=-L_o^T,&& \tilde{D}_{t,r}=D.
\end{align}
Note that the frequency-weighted observability Gramian of the pair $(\tilde{A}_{t,r},\tilde{C}_{t,r})$ is equal to $\mathscr{Q}_{s}$. Now the frequency-weighted observability Gramian $\tilde{Q}_{e}$ of the pair $(\tilde{A}_r,\tilde{C}_r)$ can be recovered from that of the pair $(\tilde{A}_{t,r},\tilde{C}_{t,r})$ using the similarity transformation matrix $\mathscr{Q}_{s}$ as $\tilde{Q}_{e}=\mathscr{Q}_{s}^{-1}\mathscr{Q}_{s}\mathscr{Q}_{s}^{-1}=\mathscr{Q}_{s}^{-1}$.\\
(ii) Consider the following Sylvester equation
\begin{align}
\tilde{A}_r^T\mathscr{Q}_{s}^{-1}\tilde{W}_b^T+\mathscr{Q}_{s}^{-1}\tilde{W}_b^T\mathscr{A}_w+\tilde{C}_r^T(\mathscr{B}_w^T\mathscr{Q}_w+\mathscr{D}_w^T\mathscr{C}_w)\nonumber\\
\mathscr{Q}_{s}^{-1}\big(-S\tilde{W}_b^T+\tilde{W}_b^T\mathscr{A}_w-L_o(\mathscr{B}_w^T\mathscr{Q}_w+\mathscr{D}_w^T\mathscr{C}_w)\big)=0.\nonumber
\end{align}
Due to uniqueness, $\mathscr{Q}_s^{-1}\tilde{W}_b^T=\tilde{\mathscr{Q}}_{12}$.\\
(iii) Consider the following Sylvester equation
\begin{align}
&\tilde{A}_r^T\tilde{Q}_e\tilde{W}_r^T+\tilde{Q}_e\tilde{W}_r^TA+\tilde{\mathscr{Q}}_{12}\mathscr{B}_wC+\tilde{C}_r^T\mathscr{B}_w^T\mathscr{Q}_{12}^T+\tilde{C}_r^T\mathscr{D}_w^T\mathscr{D}_wC\nonumber\\
&=-\tilde{Q}_eS\tilde{W}_r^T+\tilde{Q}_e\big[S\tilde{W}_r^T-\tilde{Q}_e^{-1}\tilde{\mathscr{Q}}_{12}\mathscr{B}_wC\nonumber\\
&\hspace*{0.8cm}+L_o(\mathscr{B}_w^T\mathscr{Q}_{12}+\mathscr{D}_w^T\mathscr{D}_wC)\big]+\tilde{\mathscr{Q}}_{12}\mathscr{B}_wC-\tilde{Q}_eL_o(\mathscr{B}_w^T\mathscr{Q}_{12}^T-\mathscr{D}_w^T\mathscr{D}_wC)=0.\nonumber
\end{align}
Due to uniqueness, $\tilde{Q}_e\tilde{W}_r^T=\hat{\mathscr{Q}}_{12}^T$, and thus $\tilde{Q}_e\tilde{B}_r=\hat{\mathscr{Q}}_{12}^TB$.\\
(iv) Since $\tilde{A}_r=-\mathscr{Q}_{s}S^T\mathscr{Q}_{s}^{-1}$, $\lambda_i(\tilde{A}_r)=\lambda_i(-S^T)$. Thus, $\tilde{G}_r(s)$ has poles at $-\sigma_i$.\\
\\Now recall
\begin{align}
||\mathscr{E}_o(s)||^2_{\mathcal{H}_2}=tr(B^TQ_eB)-2tr(B^T\hat{\mathscr{Q}}_{12}\tilde{B}_r)+tr(\tilde{B}_r^T\tilde{Q}_e\tilde{B}_r).\nonumber
\end{align} Since the ROM generated by O-POWI ensures that $\tilde{Q}_e\tilde{B}_r=\hat{\mathscr{Q}}_{12}^TB$,
\begin{align}
||\mathscr{E}_o(s)||^2_{\mathcal{H}_2}&=tr(B^TQ_eB)-2tr(B^T\hat{\mathscr{Q}}_{12}\tilde{B}_r)+tr(\tilde{B}_r^T\tilde{Q}_e\tilde{B}_r)\nonumber\\
&=tr\big(B^T(Q_e-\tilde{W}_r\tilde{Q}_e\tilde{W}_r^T)B\big)\nonumber\\
&=tr\big(B^T(Q_e-\tilde{W}_r\mathscr{Q}_s^{-1}\tilde{W}_r^T)B\big).\nonumber
\end{align} Thus O-POWI implicitly generates an approximation of $Q_e$, i.e., $\hat{Q}_e=\tilde{W}_r\mathscr{Q}_s^{-1}\tilde{W}_r^T$. $Q_e$ can be replaced with $\hat{Q}_e$ in FWBT to construct a ROM, and the solution of large-scale Lyapunov equation (\ref{25C}) can be avoided to save computational effort. A pseudo-code for O-POWI is given in Algorithm 2.\\
\\\textbf{Algorithm 2: O-POWI}\\
\textbf{\textit{Input:}} Original system : $(A,B,C,D)$; output weight $(\mathscr{A}_w,\mathscr{B}_w,\mathscr{C}_w,\mathscr{D}_w)$; interpolation points $(\sigma_1,\cdots,\sigma_r)$; left tangential directions $(\hat{l}_1,\cdots,\hat{l}_r)$.\\
\textbf{\textit{Output:}} ROM: $(\tilde{A}_r,\tilde{B}_r,\tilde{C}_r,\tilde{D}_r)$.\\
(i) Compute $\mathscr{Q}_w$ and $\mathscr{Q}_{12}$ from the equations (\ref{10}) and (\ref{24F}), respectively.\\
(ii) Define $\mathscr{A}_o$ and $\mathscr{C}_{\mathscr{G}}$ according to the equations (\ref{17F}) and (\ref{71F}), respectively.\\
(iii) Compute $\tilde{W}_a=\begin{bmatrix}\tilde{W}_r^T&\tilde{W}_b^T\end{bmatrix}^T$ from the equation (\ref{75F}).\\
(iv) Set $\tilde{V}_o=\tilde{W}_a$, $\tilde{W}_o=\tilde{V}_o(\tilde{V}_o^T\tilde{V}_o)^{-1}$, and compute $S$ and $L_o$ from the equations (\ref{76F}) and (\ref{77F}), respectively.\\
(v) Define $L_{o,1}$ and $L_{o,2}$ according to the equations (\ref{85F}) and (\ref{86F}), respectively.\\
(vi) Compute $\mathscr{Q}_s$ from the equation (\ref{84F}).\\
(vii) Compute the ROM from the equation (\ref{78D}).\\
\\\textbf{Remark 2:} When $\mathscr{W}(s)=I$, O-POWI becomes equivalent to PORK \citep{wolf2014h}, and when $\mathscr{W}(s)$ is an ideal (infinite-order) bandpass filter, it becomes equivalent to FLPORK \citep{zulfiqar2020frequency}.
\subsection{D-POWI}
In this subsection, we consider the FWMOR problem
\begin{align}
\underset{\substack{\tilde{G}_r(s)\\\mathscr{V}(s)\neq I,\mathscr{W}(s)\neq I}}{\text{min}}||\mathscr{W}(s)\big(G(s)-\tilde{G}_r(s)\big)\mathscr{V}(s)||^2_{\mathcal{H}_{2}}.\nonumber
\end{align}
There can be two possible ways to construct a ROM for this case by using the results of I-POWI and O-POWI. The first approach can be to use $\hat{P}_e$ and $\hat{Q}_e$ generated by I-POWI and O-POWI, respectively, for constructing a frequency-weighted balanced realization. A ROM can then be obtained by truncation. The second approach can be to use an IRKA-type framework to obtain a ROM which satisfies the following interpolation conditions
\begin{align}
\mathscr{F}[G(-\lambda_i)]b_i&=\mathscr{F}[\tilde{G}_r(-\lambda_i)]b_i,\nonumber\\
c_i^T\mathscr{G}[G(-\lambda_i)]&=c_i^T\mathscr{G}[\tilde{G}_r(-\lambda_i)].\nonumber
\end{align}
Such a ROM satisfies the conditions (\ref{49C}) and (\ref{50C}), and thus it ensures overall less $||\mathscr{W}(s)\big(G(s)-\tilde{G}_r(s)\big)\mathscr{V}(s)||_{\mathcal{H}_2}$. There are some practical limitations in achieving such a ROM using an IRKA-type framework, which we discuss after presenting the main algorithm. D-POWI closely resembles the heuristic modification of IRKA presented in \citep{zulfiqar2018weighted}, which was based on the experimental results, and the connection with the optimality conditions was left as future work. This paper fills this gap and establishes this connection.\\
\\\textbf{Algorithm 3: D-POWI}\\
\textbf{\textit{Input:}} Original system: $(A,B,C,D)$; input weight: $(\mathscr{A}_v,\mathscr{B}_v,\mathscr{C}_v,\mathscr{D}_v)$; output weight: $(\mathscr{A}_w,\mathscr{B}_w,\mathscr{C}_w,\mathscr{D}_w)$; interpolation points $(\sigma_1,\cdots,\sigma_r)$; right tangential directions $(\hat{r}_1,\cdots,\hat{r}_r)$; left tangential directions $(\hat{l}_1,\cdots,\hat{l}_r)$.\\
\textbf{\textit{Output:}} ROM ($\tilde{A}_r,\tilde{B}_r,\tilde{C}_r,\tilde{D}_r$).\\
(i) Compute $\mathscr{P}_v$ and $\mathscr{P}_{12}$ from the equations (\ref{7}) and (\ref{18C}), respectively.\\
(ii) Define $\mathscr{A}_i$ and $\mathscr{B}_{\mathscr{F}}$ according to the equations (\ref{13F}) and (\ref{34F}), respectively.\\
(iii) Compute $\mathscr{Q}_w$ and $\mathscr{Q}_{12}$ from the equations (\ref{10}) and (\ref{24F}), respectively.\\
(iv) Define $\mathscr{A}_o$ and $\mathscr{C}_{\mathscr{G}}$ according to the equations (\ref{17F}) and (\ref{71F}), respectively.\\
\textbf{while} the relative change in $\Lambda$ $>$ tolerance\\
(v)
\begin{align}
Ran\begin{bmatrix}\tilde{V}^{(a)}\\\tilde{V}^{(b)}\end{bmatrix}&=\underset {i=1,\cdots,r}{span}\{(\sigma_i I-\mathscr{A}_i)^{-1}\mathscr{B}_{\mathscr{F}} \hat{r}_i\}\nonumber\\
Ran\begin{bmatrix}\tilde{W}^{(a)}\\\tilde{W}^{(b)}\end{bmatrix}&=\underset {i=1,\cdots,r}{span}\{(\sigma_i I-\mathscr{A}_o^T)^{-1}\mathscr{C}_{\mathscr{G}}^T \hat{l}_i\}.\nonumber
\end{align}
(vi) $Ran(\tilde{V}_r)\supset Ran(\tilde{V}^{(a)})$ and $Ran(\tilde{W}_r)\supset Ran(\tilde{W}^{(a)})$.\\
(vii) $\tilde{V}_r=orth(\tilde{V}_r)$, $\tilde{W}_r=orth(\tilde{W}_r)$, $\tilde{W}_r=\tilde{W}_r(\tilde{V}_r^T\tilde{W}_r)^{-1}$.\\
(viii) $\tilde{A}_r=\tilde{W}_r^TA\tilde{V}_r$, $\tilde{B}_r=\tilde{W}_r^TB$, $\tilde{C}_r=C\tilde{V}_r$.\\
(ix) Compute the spectral factorization of $\tilde{A}_r$, i.e. $\tilde{A}_r=R\Lambda R^{-1}$, and set $\sigma_i=-\lambda_i(\tilde{A}_r)$, $\begin{bmatrix}\hat{r}_1&\cdots&\hat{r}_r\end{bmatrix}=\tilde{B}_r^*R^{-*}$, $\begin{bmatrix}\hat{l}_1&\cdots&\hat{l}_r\end{bmatrix}^*=\tilde{C}_rR$ where $\begin{bmatrix}\cdot\end{bmatrix}^*$ represents the Hermitian of the matrix.\\
\textbf{end while}\\
\\It is evident from the steps (v) and (ix) that D-POWI tends a generate a ROM that satisfies the interpolation conditions $\mathscr{F}[G(-\lambda_i)]b_i=\mathscr{F}[\tilde{G}_r(-\lambda_i)]b_i$ and $c_i^T\mathscr{G}[G(-\lambda_i)]=c_i^T\mathscr{G}[\tilde{G}_r(-\lambda_i)]$. As shown in \citep{breiten2015near} that the exact interpolation is only possible if $\mathscr{P}_{12}=\tilde{V}_r\tilde{\mathscr{P}}_{12}$ (and dually $\mathscr{Q}_{12}=\tilde{W}_r\tilde{\mathscr{Q}}_{12}$), which is generally not the case. Thus D-POWI (like NOWI) satisfies these conditions approximately, and it approximately enforces the interpolation conditions enforced by I-POWI and O-POWI. Note that I-POWI and O-POWI satisfy the respective interpolation conditions exactly, which D-POWI fails to satisfy. This is made possible by parameterizing $(\tilde{A}_r,\tilde{B}_r)$ in I-POWI and $(\tilde{A}_r,\tilde{C}_r)$ in O-POWI, and judiciously enforcing these conditions.
\subsection{Choice of the Interpolation Data}
Unlike NOWI, the poles of the ROM are known a priori in I-POWI and O-POWI. Thus the stability of the ROM is guaranteed. This information can be used to obtain a high-fidelity ROM by making a judicious choice of the interpolation points. If the ROM interpolates at the mirror images of the poles of the original system and weight that have large associated residuals, a small frequency-weighted $\mathcal{H}_2$-norm error can be achieved (as established in \citep{anic2013interpolatory} and \citep{gugercin2004krylov}). The poles of $G(s)\mathscr{V}(s)$ and $\mathscr{W}(s)G(s)$ associated with the large residuals can efficiently be captured using the eigensolver in \citep{rommes2006efficient}. I-POWI and O-POWI can then interpolate at the mirror images of these poles. For D-POWI, the mirror images of these poles can be used as a first guess. However, there is no guarantee in D-POWI that the final ROM interpolates at the mirror images of these poles.\\
I-POWI and O-POWI have an inherent property of the monotonic decay in error as the order of ROM increases, provided the new interpolation data used to construct the ROM contains the previous interpolation data. Let $\tilde{G}_{r_1}(s)$ and $\tilde{G}_{r_2}(s)$ are the $r_1$- and $r_2$-order ROMs, respectively where $r_2>r_1$. Let both $\tilde{G}_{r_1}(s)$ and $\tilde{G}_{r_2}(s)$ are obtained using I-POWI such that the interpolation data for $\tilde{G}_{r_2}(s)$ contains the interpolation data used to obtain $\tilde{G}_{r_1}(s)$. This implies that $\mathscr{F}[\tilde{G}_{r_2}(-\lambda_i)]b_i=\mathscr{F}[\tilde{G}_{r_1}(-\lambda_i)]b_i$ for $i=1,\cdots,r_1$. This further implies that
\begin{align}
||\big(\tilde{G}_{r_2}(s)-\tilde{G}_{r_1}(s)\big)\mathscr{V}&(s)||^2_{\mathcal{H}_2}=||\tilde{G}_{r_2}(s)\mathscr{V}(s)||^2_{\mathcal{H}_2}-||\tilde{G}_{r_1}(s)\mathscr{V}(s)||^2_{\mathcal{H}_2}.\nonumber
\end{align}
Thus
\begin{align}
||\tilde{G}_{r_2}(s)\mathscr{V}(s)||^2_{\mathcal{H}_2}\geq||\tilde{G}_{r_1}(s)\mathscr{V}(s)||^2_{\mathcal{H}_2}.\nonumber
\end{align}
Moreover,
\begin{align}
||\big(G(s)-\tilde{G}_{r_1}(s)\big)\mathscr{V}(s)&||^2_{\mathcal{H}_2}=||G(s)\mathscr{V}(s)||^2_{\mathcal{H}_2}-||\tilde{G}_{r_1}(s)\mathscr{V}(s)||^2_{\mathcal{H}_2},\nonumber\\
||\big(G(s)-\tilde{G}_{r_2}(s)\big)\mathscr{V}(s)&||^2_{\mathcal{H}_2}=||G(s)\mathscr{V}(s)||^2_{\mathcal{H}_2}-||\tilde{G}_{r_2}(s)\mathscr{V}(s)||^2_{\mathcal{H}_2}.\nonumber
\end{align}
Thus
\begin{align}
||\big(G(s)-\tilde{G}_{r_2}(s)\big)\mathscr{V}(s)&||^2_{\mathcal{H}_2}\leq||\big(G(s)-\tilde{G}_{r_1}(s)\big)\mathscr{V}(s)||^2_{\mathcal{H}_2}.\nonumber
\end{align}
Similarly, if $\tilde{G}_{r_1}(s)$ and $\tilde{G}_{r_2}(s)$ are obtained using O-POWI such that the interpolation data for $\tilde{G}_{r_2}(s)$ contains the interpolation data used to obtain $\tilde{G}_{r_1}(s)$, the following holds
\begin{align}
||\mathscr{W}(s)\big(G(s)-\tilde{G}_{r_2}(s)\big)&||^2_{\mathcal{H}_2}\leq||\mathscr{W}(s)\big(G(s)-\tilde{G}_{r_1}(s)\big)||^2_{\mathcal{H}_2}.\nonumber
\end{align} The error continues to decay if the interpolation data is selected like this, and a new ROM is obtained. We do not advocate that the selection of interpolation data like this is the best way as this property says nothing regarding how quickly the error decays. However, this property can potentially be exploited in an algorithm wherein the selection of interpolation data is based on some sort of optimization scheme to ensure less error. The monotonic decay of error in I-POWI and O-POWI can be exploited in such a scheme to obtain a high-fidelity ROM. This is a potential direction for future work.
\section{Computational Cost}
In the projection-based MOR algorithms, the computation of the reduction subspaces dominates the overall computational cost. Moreover, these are computed by solving some linear matrix equations of the following form
\begin{align}
KX+XL+MN=0.\label{R}
\end{align}
The rule of thumb is that the computational cost of equation (\ref{R}) is high if $K$, $L$, $M$, and $N$ are all large-scale matrices. The balancing-based methods fall in this category. The computational cost of equation (\ref{R}) is low if $K$ and $M$ are large-scale matrices, but $L$ and $N$ are small-scale matrices. The computational cost is even less if $K$ and $M$ are sparse matrices. The Krylov subspace-based methods and the low-rank versions of balancing-based methods fall in this category. The reduction subspaces in I-POWI, O-POWI, and D-POWI can be obtained by solving the linear matrix equation of the form (\ref{R}), wherein $L$ and $N$ are small-scale matrices. Further, I-POWI and O-POWI generate the ROM in a single-run, i.e., the equation (\ref{R}) is solved once. NOWI and DOWI solve two such equations once in every iteration. Thus I-POWI and O-POWI have a computational cost, which is a fraction of that of NOWI and D-POWI. Moreover, $L$ and $N$ are large-scale in FWBT that makes it computationally infeasible in the large-scale setting.

We now analyze the computational cost in detail. We first divide the FWMOR problem into two categories before discussing the computational cost associated with each one. The first case is when $n_v<<n$ and $n_w<<n$, i.e., the weights $\mathscr{V}(s)$ and $\mathscr{W}(s)$ are small-scale systems, and $G(s)$ is a large-scale system. In this case, the computational cost of FWBT is dominated by the solutions of large-scale Lyapunov equations (\ref{17C}) and (\ref{25C}), i.e., the computation of $P_e$ and $Q_e$. This is because the computation of $\mathscr{P}_v$ and $\mathscr{Q}_w$ involve only small-scale Lyapunov equations, which can be solved with minimal computational effort. Also, the computation of $\mathscr{P}_{12}$ and $\mathscr{Q}_{12}$ is dominated by the small-scale $\mathscr{A}_v$ and $\mathscr{A}_w$, and hence, these can also be solved within admissible time \citep{bennersparse}. Note that NOWI and POWI avoid the computation of the large-scale Lyapunov equations (\ref{17C}) and (\ref{25C}), and thus their computational cost is dominated by the rational Krylov subspaces $\tilde{V}_r$ and $\tilde{W}_r$. The rational Krylov subspaces can be computed within the admissible time for large-scale systems, and this property is the main advantage of the rational Krylov subspace-based MOR algorithms over balanced realization-based MOR algorithms. Note that I-POWI and O-POWI are iteration-free, and thus their computational cost is a fraction of that of NOWI. However, D-POWI is an iterative algorithm similar to NOWI, and hence, its computational cost is comparable to that of NOWI. Moreover, POWI also provides approximations of $P_e$ and $Q_e$, which can make FWBT computationally as efficient as POWI. We refer the readers to the section $5$ of \citep{breiten2015near} for the computational cost comparison of NOWI and FWBT.

Next, we consider the case wherein $n_v$ and $n_w$ is comparable to $n$, and $G(s)$ is large-scale. In this scenario, the computation of $\mathscr{P}_v$, $\mathscr{Q}_w$, $\mathscr{P}_{12}$, and $\mathscr{Q}_{12}$ also becomes a computational challenge. FWBT, in addition, requires the computation of $P_e$ and $Q_e$. Therefore, FWBT is computationally more expensive as compared to NOWI and POWI. If the low-rank approximations of $\mathscr{P}_v$, $\mathscr{Q}_w$, $\mathscr{P}_{12}$, and $\mathscr{Q}_{12}$ are obtained (for instance by using the algorithms in \citep{ahmad2010krylov}), NOWI and POWI can be executed within the admissible time. However, FWBT remains computationally expensive due to the computation of $P_e$ and $Q_e$. FWBT can be executed within the admissible time if $P_e$ and $Q_e$ are replaced with $\hat{P}_e$ and $\hat{Q}_e$ generated by POWI. To best of authors' knowledge, POWI is the first method of its kind in the literature so far for approximating large-scale frequency-weighted system Gramians; however, there exist several methods for computing low-rank approximation of large-scale Lyapunov equations \citep{benner2013numerical}, which can be used to approximate the equations (\ref{17C}) and (\ref{25C}).
\section{Numerical Validation}
In this section, we validate the theory developed in the previous section, and we test the performance of the developed algorithms. We compare our algorithms with FWBT and NOWI, which are the gold standards of the frequency-weighted MOR algorithms. We present three numerical examples for this purpose. The first example is an illustrative example that is presented to aid a convenient repeatability for the readers to verify all the important properties of the algorithms mentioned in the last section. The second example is a model reduction problem wherein it is shown the ROMs exhibit superior accuracy in the frequency region emphasized by the frequency weight. The third example is a controller reduction problem wherein the order of the controller is reduced using the proposed algorithms. In all the examples, we solve the Lyapunov and Sylvester equations exactly using MATLAB's \textit{``lyap"} command. All the experiments are performed on a Core i7 computer with $16$GB RAM using MATLAB $2016$.
\\\\\textbf{Illustrative Example}\\
Consider a $3^{rd}$ order MIMO system with $3$-inputs and $2$-outputs with the following state-space realization
\begin{align}
A&=\begin{bmatrix} -0.4727  &  0.1422  & -2.9044\\
    0.3754 &  -0.9764  & -1.1972\\
    2.8836  &  1.2466  & -0.3644\end{bmatrix},\hspace*{2mm}B=\begin{bmatrix} 0     &    0  &  0.7916\\
         0   & 1.5677  & -0.0930\\
   -2.7018  &       0   & -0.3802\end{bmatrix},\nonumber\\
C&=\begin{bmatrix}0.6959  & -0.2684  & -0.5393\\
         0   & 1.4370  & -0.4301\end{bmatrix},\hspace*{5.2mm}D=\begin{bmatrix}  0   &      0  & -2.4207\\
   -0.9021  & -1.6833  &       0\end{bmatrix}.\nonumber
\end{align}
Let the input frequency weight has the following state-space realization
\begin{align}
\mathscr{A}_v=&\begin{bmatrix}-0.9452 &   0.0546\\
    0.0546 &  -1.0319\end{bmatrix},\hspace*{2mm}\mathscr{B}_v=\begin{bmatrix}
    0.3656    &     0  &  0.5451\\
   -0.8849 &  -2.6384   & 1.0780\end{bmatrix},\nonumber\\
\mathscr{C}_v=&\begin{bmatrix} 0 &  -1.3113\\
    2.3793 &  -0.1457\\
   -0.6410  &  0.1058\end{bmatrix},\hspace*{2mm}\mathscr{D}_v=\begin{bmatrix} 0     &    0  &  0.7236\\
         0  & -0.5867     &    0\\
   -0.7636  &       0      &   0\end{bmatrix}.\nonumber
\end{align}
Consider the following interpolation data for I-POWI: $\sigma_1=1$ and $\hat{r}_1=\begin{bmatrix}1&1&1\end{bmatrix}^T$.
Then the ROM generated by I-POWI is given by
\begin{align}
\tilde{A}_r&=\begin{bmatrix}-1\end{bmatrix},\hspace*{2mm}\tilde{B}_r=\begin{bmatrix}-1.2839 &  -1.2839 &  -1.2839
\end{bmatrix},\hspace*{2mm}\tilde{C}_r=\begin{bmatrix}-0.3481& -0.6602
\end{bmatrix}^T.\nonumber
\end{align}
The optimality condition can be verified as $
\tilde{C}_r\tilde{P}_e=C\hat{\mathscr{P}}_{12}=\begin{bmatrix}-1.3912\\-2.6384\end{bmatrix}$.

Let the output frequency weight has the following state-space realization
\begin{align}
\mathscr{A}_w=&\begin{bmatrix}-1.6503   & 1.6670\\
    1.6670  & -2.0860\end{bmatrix},\hspace*{2mm}\mathscr{B}_w=\begin{bmatrix}0.1897  & -0.4772\\
   -0.4555  & -0.2561\end{bmatrix},\nonumber\\
\mathscr{C}_w=&\begin{bmatrix}0.7987   & 2.0373\\
         0 &  -0.3397
\end{bmatrix},\hspace*{4.7mm}\mathscr{D}_w=\begin{bmatrix}0 &   0.2353\\
    0.5445       &  0\end{bmatrix}.\nonumber
\end{align}
Consider the following interpolation data for O-POWI: $\sigma_1=1$ and $\hat{l}_1=\begin{bmatrix}1&1\end{bmatrix}$.
Then the ROM generated by O-POWI is given by
\begin{align}
\tilde{A}_r&=\begin{bmatrix} -1
\end{bmatrix},\hspace*{2mm}\tilde{B}_r=\begin{bmatrix}0.3875  &  0.6217  & -0.1832 \end{bmatrix},\hspace*{2mm}\tilde{C}_r=\begin{bmatrix}1.5952& 1.5952 \end{bmatrix}^T.\nonumber
\end{align}
The optimality conditions can be verified as
\begin{align}
\tilde{Q}_e\tilde{B}_r=\hat{\mathscr{Q}}_{12}^TB=\begin{bmatrix}4.1096  &  6.5939 &  -1.9429\end{bmatrix}.\nonumber
\end{align}

\begin{figure}[!h]
\centering
\includegraphics[width=9cm]{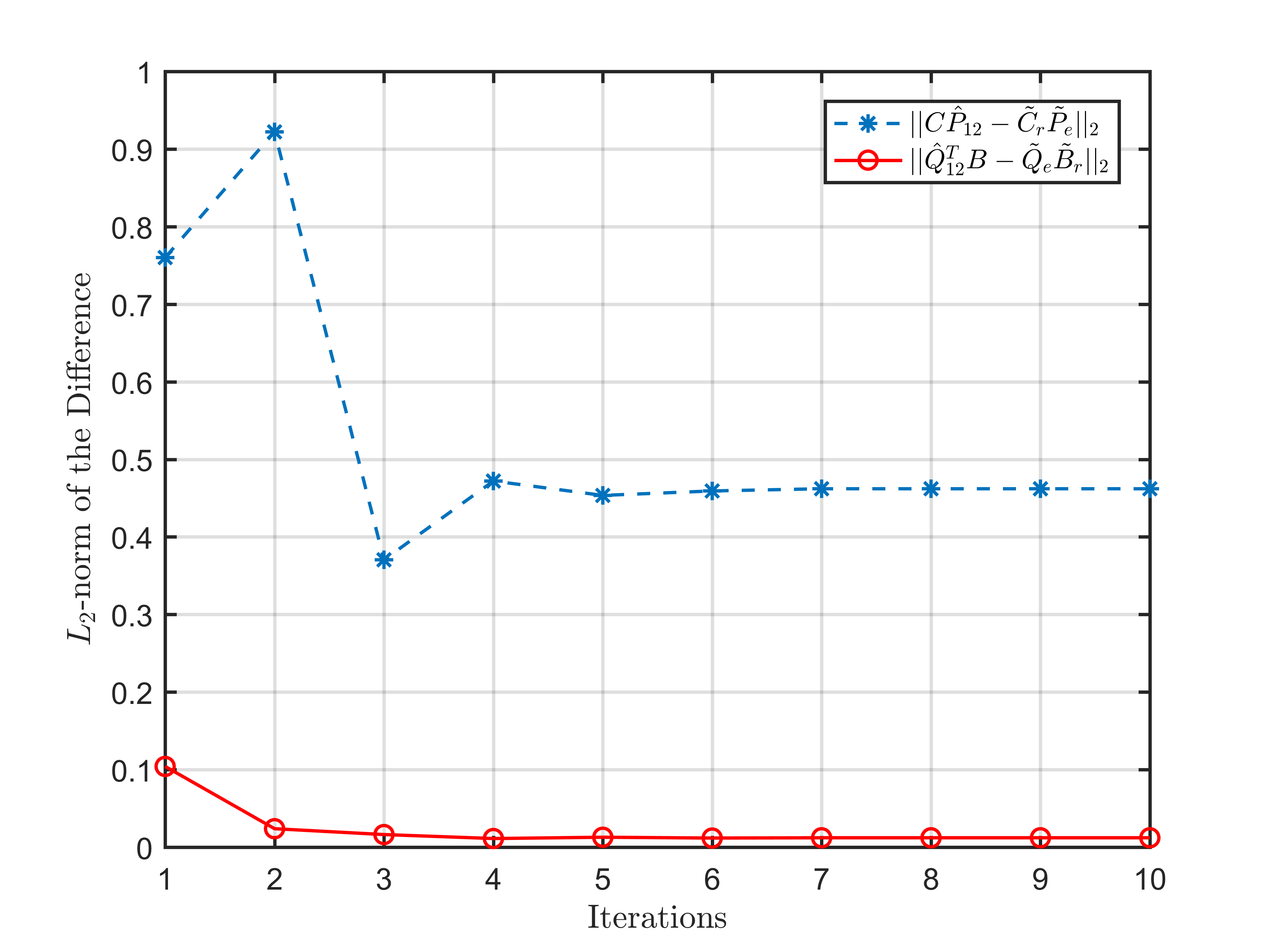}
\caption{Deviation from the optimality conditions}\label{fig1}
\end{figure}

\begin{figure}[!h]
\centering
\includegraphics[width=9cm]{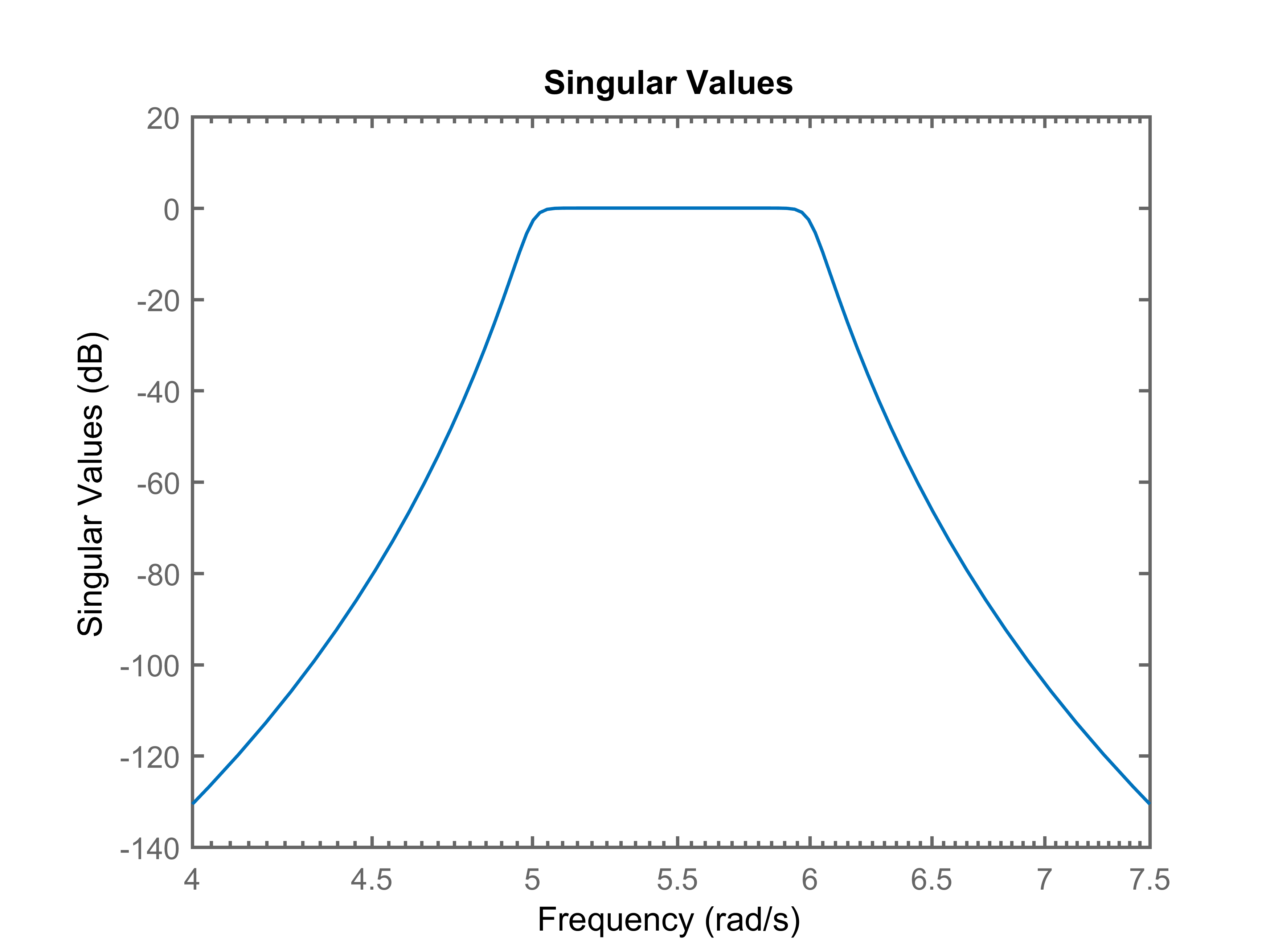}
\caption{The sigma plot of the bandpass filter}\label{fig2}
\end{figure}

\begin{figure}[!h]
\centering
\includegraphics[width=9cm]{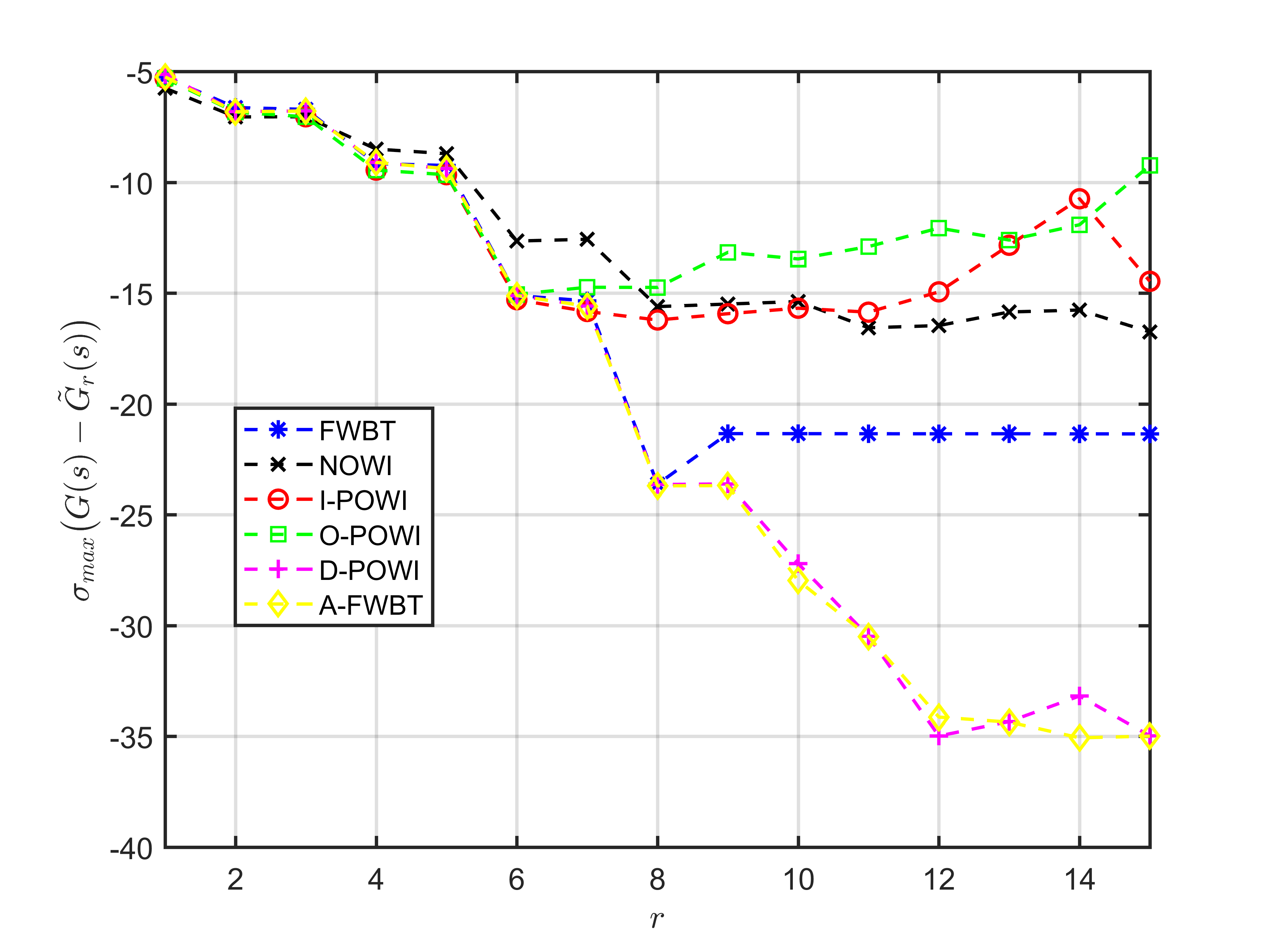}
\caption{Maximum singular value of error within $\{5,6\}$ rad/sec}\label{fig3}
\end{figure}

Next, we obtain the first-order ROM using D-POWI. We initialize D-POWI with the following interpolation data: $\sigma_1=1$, $\hat{r}_1=\begin{bmatrix}1&1&1\end{bmatrix}^T$, and $\hat{l}_1=\begin{bmatrix}1&1\end{bmatrix}$. We know that the ROM generated by D-POWI does not achieve the optimality conditions $C\hat{\mathscr{P}}_{12}=\tilde{C}_r\tilde{P}_e$ and $\hat{\mathscr{Q}}_{12}^TB=\tilde{Q}_e\tilde{B}_r$. Therefore, we investigate the differences $||C\hat{\mathscr{P}}_{12}-\tilde{C}_r\tilde{P}_e||_2$ and $||\hat{\mathscr{Q}}_{12}^TB-\tilde{Q}_e\tilde{B}_r||_2$ as the iterations of D-POWI progress. The ROM obtained using D-POWI is given by
\begin{align}
\tilde{A}_r=\begin{bmatrix}-5.8318\end{bmatrix},\hspace*{2mm} \tilde{B}_r=\begin{bmatrix}-4.2486&-5.5076&1.4409\end{bmatrix},\hspace*{2mm}
\tilde{C}_r=\begin{bmatrix}-0.5643\\-0.9750\end{bmatrix}.\nonumber
\end{align}
It can be seen from FIG. \ref{fig1} that the differences in $||C\hat{\mathscr{P}}_{12}-\tilde{C}_r\tilde{P}_e||_2$ and $||\hat{\mathscr{Q}}_{12}^TB-\tilde{Q}_e\tilde{B}_r||_2$ are stagnated after a few iterations and are greater than $0$.
Thus D-POWI generates a ROM which approximately satisfies $C\hat{\mathscr{P}}_{12}=\tilde{C}_r\tilde{P}_e$ and $\hat{\mathscr{Q}}_{12}^TB=\tilde{Q}_e\tilde{B}_r$. The differences $C\hat{\mathscr{P}}_{12}-\tilde{C}_r\tilde{P}_e$ and $\hat{\mathscr{Q}}_{12}^TB-\tilde{Q}_e\tilde{B}_r$ of the final ROM are given by
\begin{align}
C\hat{\mathscr{P}}_{12}-\tilde{C}_r\tilde{P}_e=\begin{bmatrix}-0.2306 \\-0.4004\end{bmatrix},\hspace*{2mm}\hat{\mathscr{Q}}_{12}^TB-\tilde{Q}_e\tilde{B}_r=\begin{bmatrix}-0.0071 &  -0.0095  &  0.0026\end{bmatrix}.\nonumber
\end{align}
\\\textbf{Model Reduction: Los Angeles Building}\\
In this experiment, we consider the $48^{th}$-order SISO model of the Los Angeles building from the benchmark collection of \citep{chahlaoui2005benchmark}. We design a $24^{th}$-order bandpass filter with a passband of $\{5,6\}$ rad/sec using MATLAB's ``\textit{butter}" command. The frequency-domain plot of the filter is shown in FIG. \ref{fig2}. Clearly, it emphasizes the frequency region $\{5,6\}$ rad/sec, and thus we use this filter as the input and output frequency weights to obtain superior accuracy within $\{5,6\}$ rad/sec.
We obtain ROMs ranging from order $1-15$ using FWBT, NOWI, I-POWI, O-POWI, and D-POWI. NOWI and D-POWI are initialized randomly, while the final interpolation data of D-POWI is used for I-POWI and O-POWI. This helps us to investigate the effect on satisfying a subset of the optimality conditions on the accuracy of the ROM. The approximate frequency-weighted Gramians obtained from I-POWI and O-POWI are used to implement approximate FWBT (A-FWBT). To compare the ROMs generated by these different frequency-weighted MOR cases, we use the maximum singular value of the error transfer within $\{5,6\}$ rad/sec as a common yardstick. This selection is based on the fact that all these techniques tend to ensure superior accuracy within $\{5,6\}$ rad/sec due to the selection of the bandpass filter as the frequency weights. The singular values of the error within $\{5,6\}$ rad/sec are computed using MATLAB's \textit{``sigma"} command. The maximum singular values of $G(s)-\tilde{G}_r(s)$ within $\{5,6\}$ rad/sec are plotted in FIG. \ref{fig3} on a logarithmic scale.
 The accuracy of all the techniques is comparable for the order $1-7$. Thereafter, the effect of double-sided weights in FWBT, D-POWI, and A-FWBT becomes apparent as the error drops significantly than the single-sided NOWI, I-POWI, and O-POWI.\\
\\\textbf{Controller Reduction: International Space Station}\\
In this experiment, we consider the $270^{th}$-order MIMO model of the international space station from the benchmark collection of \citep{chahlaoui2005benchmark} as the plant $P(s)$. We design a $260^{th}$-order $\mathcal{H}_{\infty}$-controller using the Glover-McFarlane loop-shaping method \citep{mcfarlane1992loop}. We use $W_i(s)=\frac{10}{s+1}I_{3\times 3}$ as the shaping filter to ensure that there is a high loop-gain at the low-frequencies and a low loop-gain at high frequencies. We use MATLAB's ``\textit{ncfsyn}" command to design the controller $K(s)$, which is a stabilizing controller for $P(s)$. We know from \citep{enns1985} that $K_r(s)$ is also a stabilizing controller for $P(s)$ if $K_r(s)$ has same number of poles in the right half of $s$-plane as that of $K(s)$, and
\begin{figure}[!h]
\centering
\includegraphics[width=9cm]{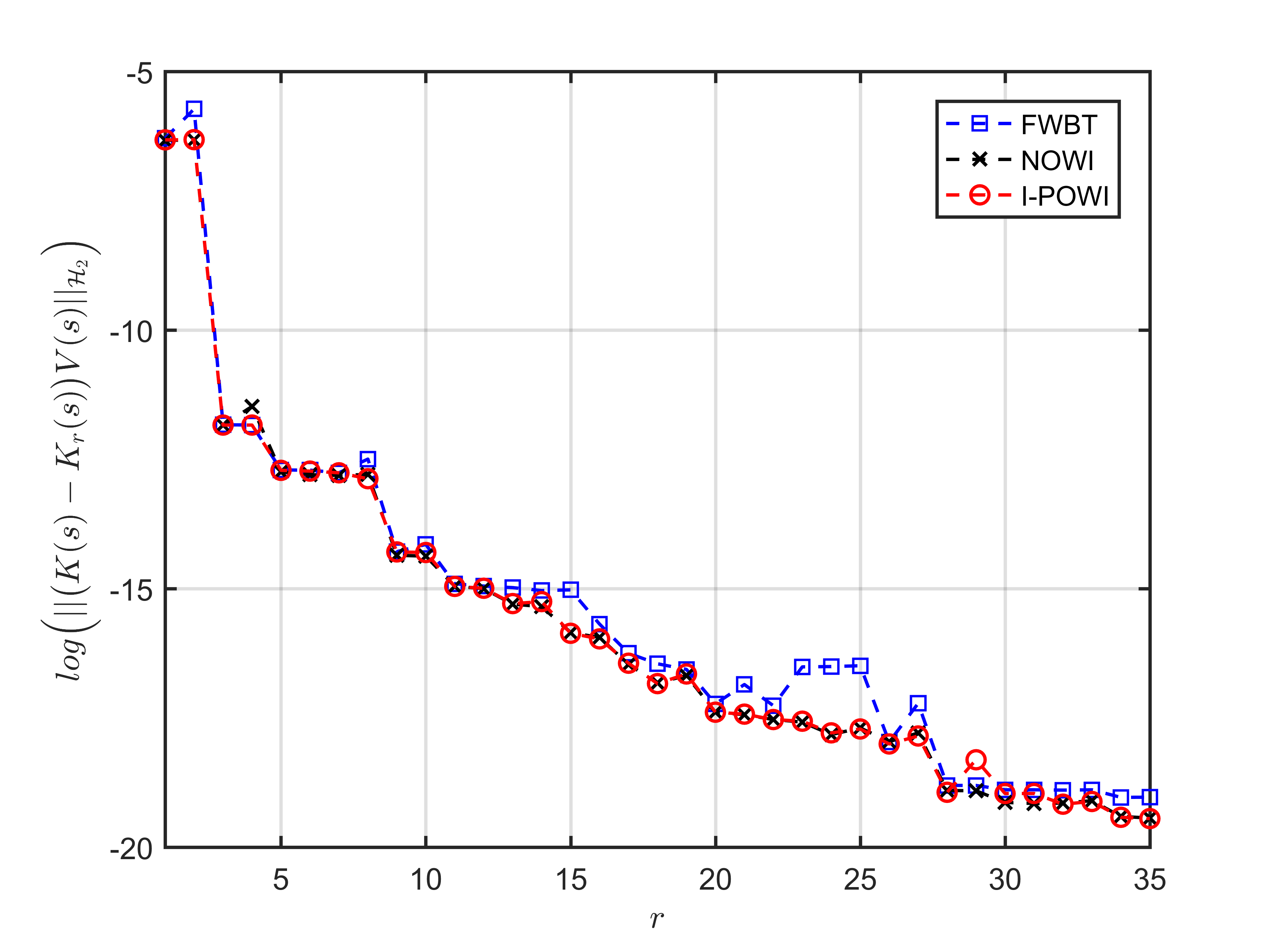}
\caption{Frequency-weighted $\mathcal{H}_2$-norm error}\label{fig4}
\end{figure}
\begin{figure}[!h]
\centering
\includegraphics[width=9cm]{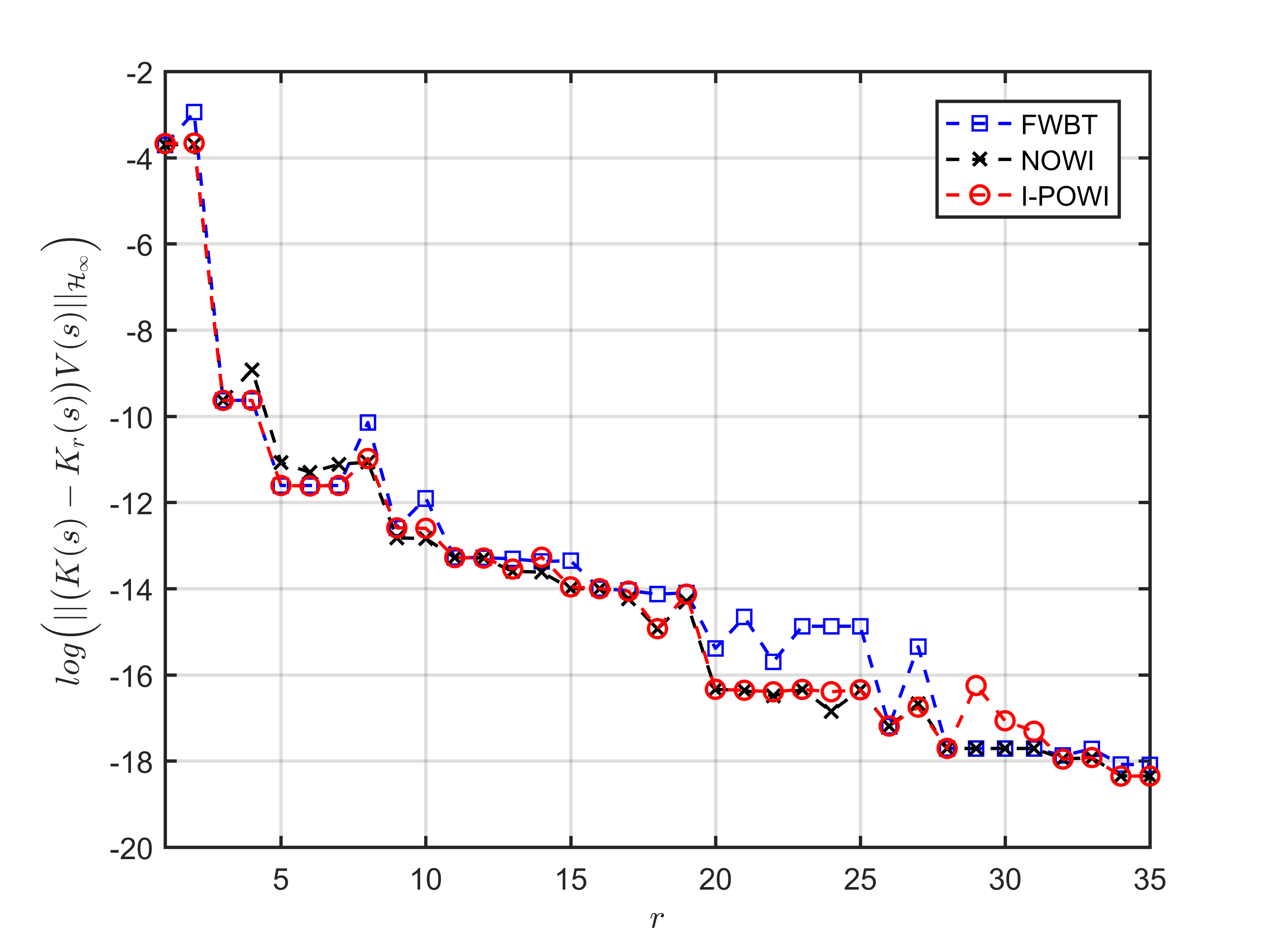}
\caption{Frequency-weighted $\mathcal{H}_\infty$-norm error}\label{fig5}
\end{figure}
\begin{figure}[!h]
\centering
\includegraphics[width=9cm]{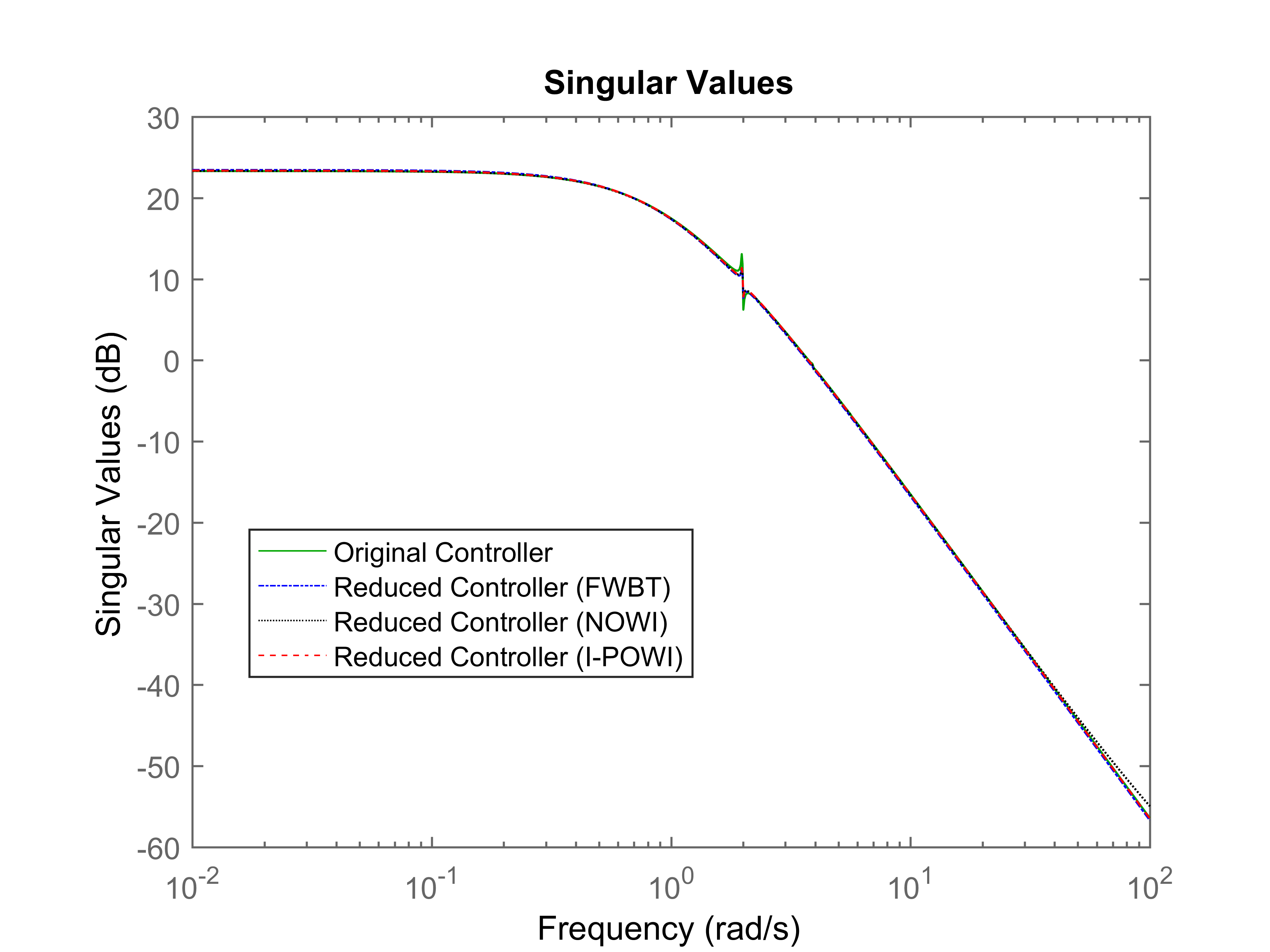}
\caption{Loop-shape (Output-1) achieved using $K(s)$ and $K_r(s)$}\label{fig6}
\end{figure}
\begin{figure}[!h]
\centering
\includegraphics[width=9cm]{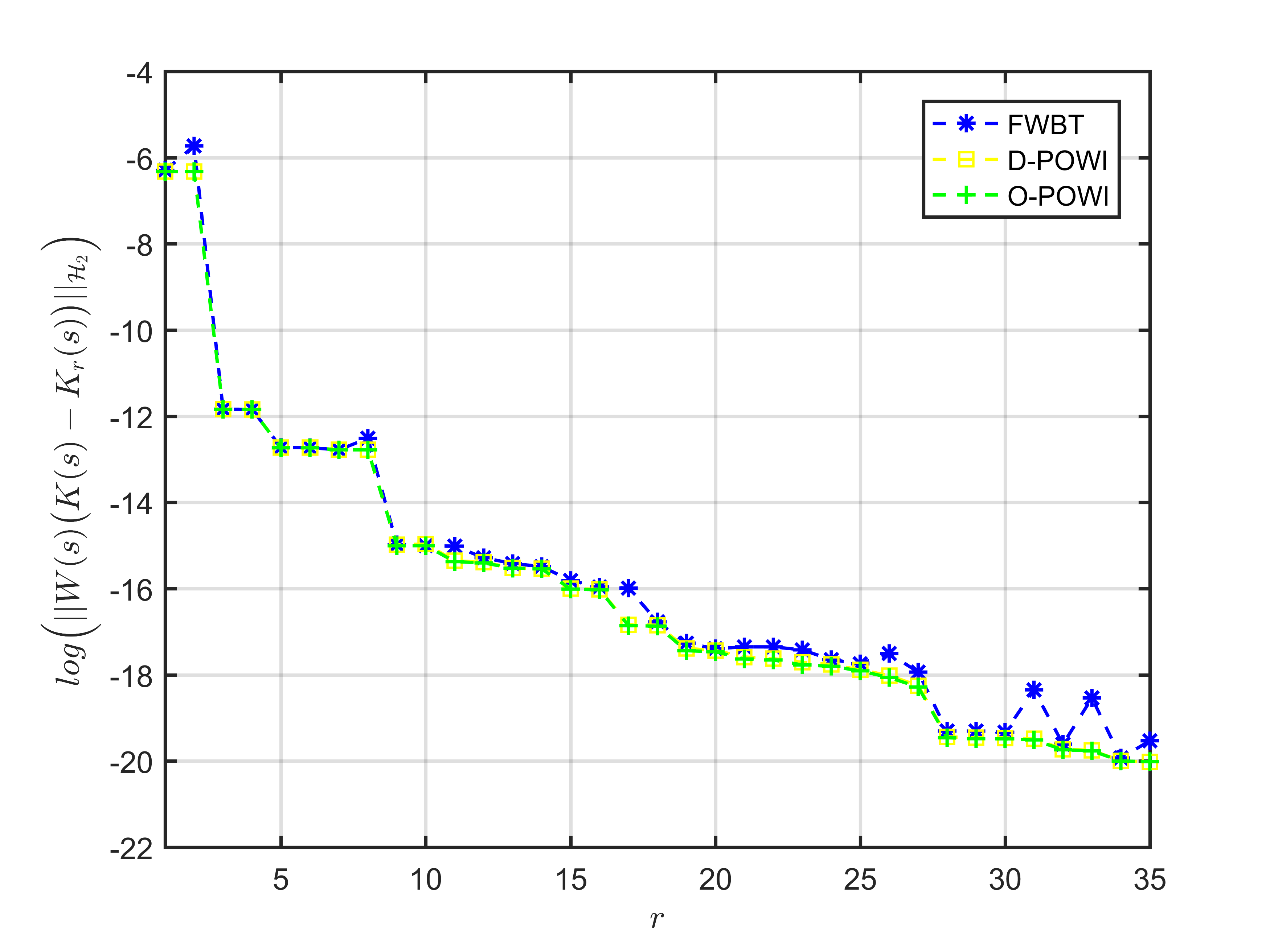}
\caption{Frequency-weighted $\mathcal{H}_2$-norm error}\label{fig7}
\end{figure}
\begin{figure}[!h]
\centering
\includegraphics[width=9cm]{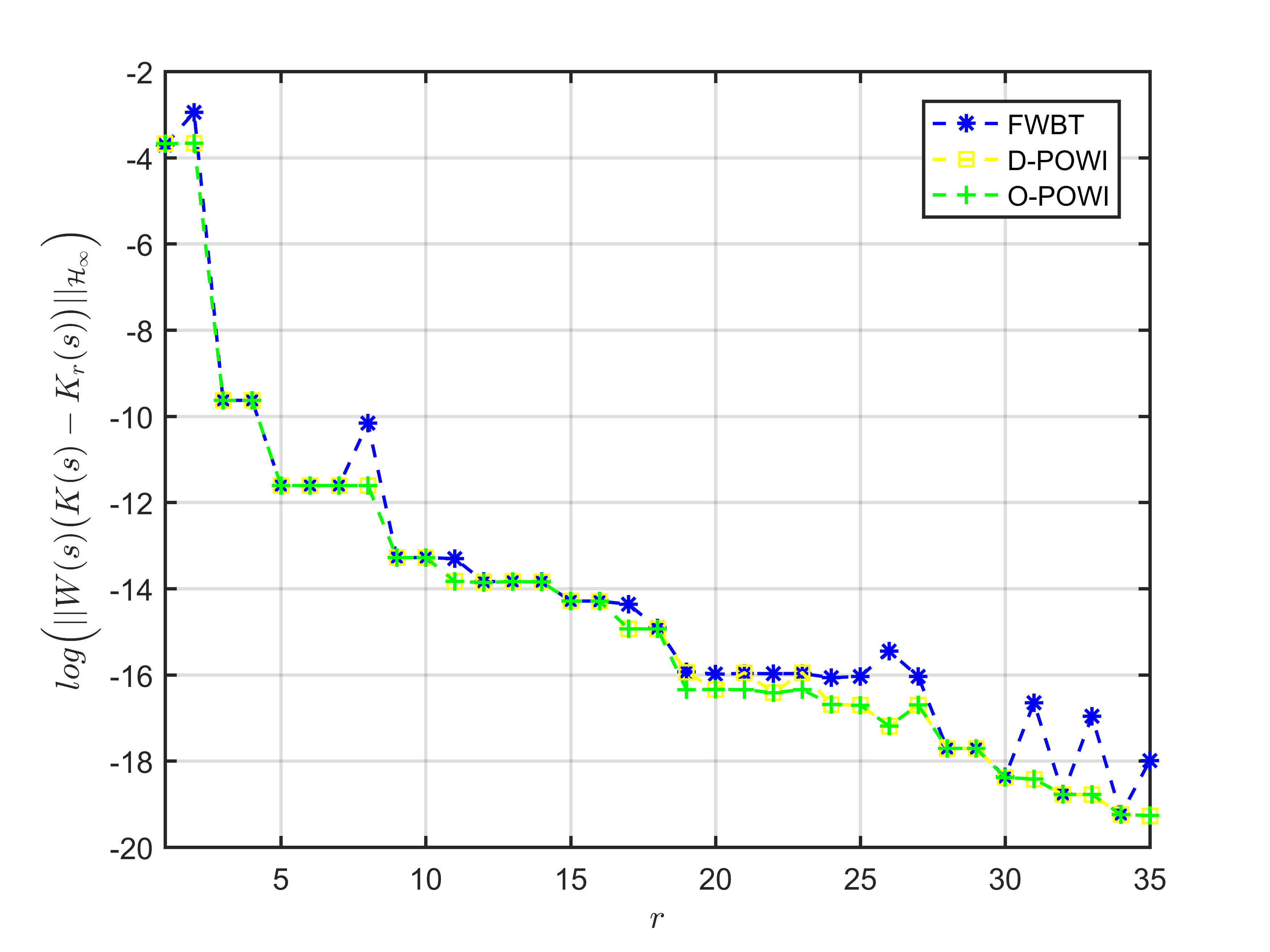}
\caption{Frequency-weighted $\mathcal{H}_\infty$-norm error}\label{fig8}
\end{figure}
\begin{figure}[!h]
\centering
\includegraphics[width=9cm]{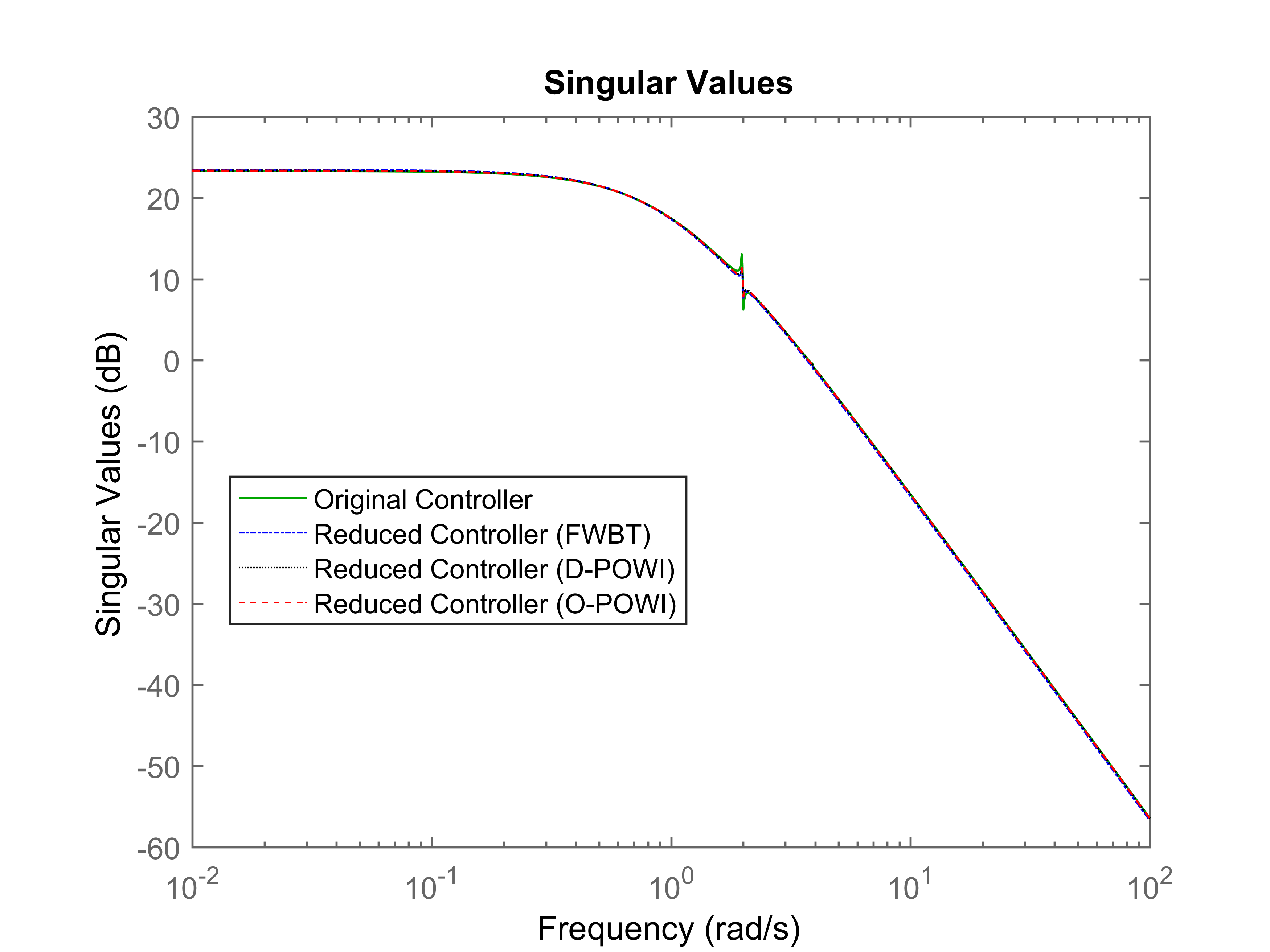}
\caption{Loop-shape (Output-1) achieved using $K(s)$ and $K_r(s)$}\label{fig9}
\end{figure}
\begin{align}
||(K(s)-K_r(s))V(s)||_{\mathcal{H}_\infty}&<1\label{eq:65}\textnormal{ or}\\
||W(s)(K(s)-K_r(s))||_{\mathcal{H}_\infty}&<1\label{eq:66}
\end{align} where
\begin{align}
V(s)=P(s)(I+K(s)P(s))^{-1}\textnormal{ and } W(s)=(I+P(s)K(s))^{-1}P(s).\nonumber
\end{align}
For the condition (\ref{eq:65}), we obtain ROMs of $K(s)$ ranging from $1-35$ using FWBT, NOWI, and I-POWI. For the input-sided case, D-POWI is quite similar to NOWI, and thus we refrain from showing results of D-POWI in FIG. \ref{fig4}-\ref{fig6} for clarity. We initialize NOWI randomly and use its final interpolation data for I-POWI. This helps us to investigate the effect on satisfying a subset of the optimality conditions on the accuracy of the ROM. The frequency-weighted $\mathcal{H}_2$- and $\mathcal{H}_{\infty}$-norms errors are plotted in FIG. \ref{fig4} and \ref{fig5}, respectively on a logarithmic scale. All the ROMs ensure the condition (\ref{eq:65}), and all are stabilizing controllers for $P(s)$. The accuracy of all the techniques is comparable in this experiment. The loop-shape achieved using $1^{st}$ ROMs is compared in FIG. \ref{fig6}. It can be noted the loop-shape is almost indistinguishable even with the first-order ROM.

For the condition (\ref{eq:66}), we obtain ROMs of $K(s)$ ranging from $1-35$ using FWBT, D-POWI, and O-POWI. We initialize D-POWI randomly and use its final interpolation data for O-POWI. The frequency-weighted $\mathcal{H}_2$- and $\mathcal{H}_{\infty}$-norms errors are plotted in FIG. \ref{fig7} and \ref{fig8}, respectively on a logarithmic scale. All the ROMs ensure the condition (\ref{eq:66}), and all are stabilizing controllers for $P(s)$. The accuracy of all the techniques is comparable in this experiment. The loop-shape achieved using $1^{st}$ ROMs is compared in FIG. \ref{fig9}, and it can be seen that it is indistinguishable.

\textbf{Discussion:} It is well-known in the literature that FWBT is a gold standard for the approximation error in the $\mathcal{H}_\infty$-norm sense. From the $\mathcal{H}_2$-norm sense, it is natural to consider NOWI (and DOWI for the double-sided case) as the gold standard  because it constructs a suboptimal ROM. FIG. \ref{fig3}, \ref{fig4}, \ref{fig5}, \ref{fig7}, and \ref{fig8} show (on a logarithmic scale) that the approximation accuracy of the proposed algorithms, i.e., I-POWI, O-POWI, and A-FLBT, is very close to the gold standards, i.e., FWBT and NOWI.
\section{Conclusion}
The problem of frequency-weighted $\mathcal{H}_2$-pseudo-optimal MOR has been addressed
by presenting three algorithms, which construct reduced order models that satisfy the associated optimality conditions. The proposed algorithms compare well in approximation accuracy with the existing techniques. Moreover, these algorithms also have advantageous features like stability preservation and computational efficiency, which give them superiority over the existing techniques. We give detailed proofs of the mathematical properties of our algorithms. The numerical experiments support the results presented in the paper.
\section*{Funding}
This work is supported by the National Natural Science Foundation of China under Grant (No. $61873336$, $61873335$), and supported in part by $111$ Project (No. D$18003$). M. I. Ahmad is supported by the Higher Education Commission of Pakistan under the National Research Program for Universities Project ID $10176$.
%%%%%%%%%%%%%figure style
%%%%%%%%%%%%%%%%%%%%%%%%%%%%%%

\vspace*{6pt}
%%%%%%%%%%%%%%%%%bibliography style
%\bibliographystyle{agsm}
%\bibliography{cst}

%%%%%%%%%%%%%%%%%%%%%%%%
%%%%%%%%%%%%%%%%%%%%%%%%%%%%%%%%%%%

\end{document}